\documentclass[aps,prb,twocolumn,groupedaddress,amsmath]{revtex4-2}

\usepackage{graphicx}
\usepackage{amssymb}
\usepackage{float}
\usepackage{dcolumn}
\usepackage{subfigure}
\usepackage{threeparttable}

\usepackage{multirow}
\DeclareMathAlphabet{\mathscrbf}{OMS}{mdugm}{b}{n}
\usepackage[usenames,dvipsnames]{color}
\begin{document}

\newcommand{\intkspa}{\int \frac{\rmd^D k}{(2\pi)^D}}
\newcommand{\vn}[1]{{\boldsymbol{#1}}}
\newcommand{\polarivec}{\hat{\vn{e}}_{\lambda}}
\newcommand{\crea}[1]{{c_{#1}^{\dagger}}}
\newcommand{\annihi}[1]{{c_{#1}^{\phantom{\dagger}}}}
\newcommand{\vht}[1]{{\boldsymbol{#1}}}
\newcommand{\matn}[1]{{\bf{#1}}}
\newcommand{\matnht}[1]{{\boldsymbol{#1}}}
\newcommand{\bege}{\begin{equation}}
\newcommand{\ee}{\end{equation}}
\newcommand{\bal}{\begin{aligned}}
\newcommand{\defbar}{\overline}
\newcommand{\SM}{\scriptstyle}
\newcommand{\eal}{\end{aligned}}
\newcommand{\torkance}{t}
\newcommand{\rmd}{{\rm d}}
\newcommand{\rme}{{\rm e}}
\newcommand{\udot}{\overset{.}{u}}
\newcommand{\exponential}[1]{{\exp(#1)}}
\newcommand{\phandot}[1]{\overset{\phantom{.}}{#1}}
\newcommand{\phandag}{\phantom{\dagger}}
\newcommand{\Trace}{\text{Tr}}
\newcommand{\Bxc}{\Omega}
\newcommand{\gret}{G^{\rm R}}
\newcommand{\gadv}{G^{\rm A}}
\newcommand{\gmat}{G^{\rm M}}
\newcommand{\gles}{G^{<}}
\newcommand{\ghat}{\hat{G}}
\newcommand{\sigmahat}{\hat{\Sigma}}
\newcommand{\glesone}{G^{<,{\rm I}}}
\newcommand{\glestwo}{G^{<,{\rm II}}}
\newcommand{\glesthree}{G^{<,{\rm III}}}
\newcommand{\magdir}{\hat{\vn{n}}}
\newcommand{\sigmaret}{\Sigma^{\rm R}}
\newcommand{\sigmales}{\Sigma^{<}}
\newcommand{\sigmalesone}{\Sigma^{<,{\rm I}}}
\newcommand{\sigmalestwo}{\Sigma^{<,{\rm II}}}
\newcommand{\sigmalesthree}{\Sigma^{<,{\rm III}}}
\newcommand{\sigmaadv}{\Sigma^{A}}

\newcommand{\psibn}{\Psi_{\vn{k}n}}
\newcommand{\psibm}{\Psi_{\vn{k}m}}

\setcounter{secnumdepth}{2}
\title{Theory of unidirectional magnetoresistance and nonlinear Hall effect}
\author{Frank Freimuth$^{1,2}$}
\email[Corresp.~author:~]{f.freimuth@fz-juelich.de}
\author{Stefan Bl\"ugel$^{1}$}
\author{Yuriy Mokrousov$^{1,2}$}
\affiliation{$^{1}$Peter Gr\"unberg Institut and Institute for Advanced Simulation,
Forschungszentrum J\"ulich and JARA, 52425 J\"ulich, Germany}
\affiliation{$^2$Institute of Physics, Johannes Gutenberg University Mainz, 55099 Mainz, Germany
}
\begin{abstract}
We study the unidirectional magnetoresistance (UMR) and the nonlinear 
Hall effect (NLHE) in the ferromagnetic Rashba model. For this 
purpose we derive expressions to describe the response of the
electric current quadratic in the applied electric field. We compare
two different formalisms, namely the standard Keldysh nonequilibrium 
formalism and the Moyal-Keldysh formalism, to derive the nonlinear
conductivities of UMR and NLHE. We find that both formalisms lead to identical
numerical results when applied to the ferromagnetic Rashba model.
The UMR and the NLHE nonlinear conductivities tend to be comparable in
magnitude according to our calculations. Additionally, their
dependencies
on the Rashba parameter and on the quasiparticle broadening are
similar.
The nonlinear zero-frequency response
considered
here is several orders of magnitude higher than the one at
optical frequencies that describes the photocurrent generation in the
ferromagnetic Rashba model.
Additionally, we compare our Keldysh nonequilibrium expression
in the independent-particle approximation to literature
expressions
of the UMR that have been obtained within the constant relaxation 
time approximation of the Boltzmann formalism. We find that both
formalisms converge to the same analytical formula in the limit of
infinite relaxation time. However, remarkably, we find that the 
Boltzmann result does not correspond to the intraband term of the
Keldysh expression. Instead, the Boltzmann result corresponds to
the sum of the intraband term and an interband term that can be
brought into the form of an effective intraband term due to the
f-sum rule.
\end{abstract}

\maketitle
\renewcommand{\arraystretch}{1.3}
\section{Introduction}

In magnetic bilayers such as Co/Pt, which are composed of a
ferromagnetic layer and a heavy metal layer, 
a change in the longitudinal resistance is
observed when either the applied in-plane current
or the magnetization is 
reversed~\cite{usmr_bilayers_avci,unidirectiona_Avci,shmr_metallic_bilayers_hayashi}.
This so-called unidirectional magnetoresistance (UMR)
is proportional to $(\vn{j}\times \hat{\vn{e}}_{z})\cdot\hat{\vn{M}}$,
where $\vn{j}$, $\hat{\vn{M}}$, 
and $\hat{\vn{e}}_{z}$ denote
the electric current, the magnetization direction,
and the unit vector along the bilayer interface normal, respectively.
UMR is a nonlinear magnetoresistance, because the corresponding
voltage is quadratic in the applied electric current.
Therefore, UMR generates a 2nd harmonic voltage when an
a.c.\ current is applied.
UMR can be used 
to detect 180$^{\circ}$ magnetization 
reversal~\cite{electrical_detection_magnetization_reversal,multi_state_memory_device_usmr}
and 
to realize reversible diodes~\cite{unidirectional_theory}.
Using UMR the four different magnetic states that may be realized 
in ferromagnet/nonmagnet/ferromagnet 
trilayers may be differentiated~\cite{multi_state_memory_device_usmr}.

One contribution to the UMR arises from spin accumulation
in the ferromagnetic layer, which modifies the electrical conductivity
when the mobility is spin-dependent~\cite{unidirectional_theory}.
The spin accumulation itself may arise from the spin Hall effect (SHE) in
the heavy metal, which injects spin current into the ferromagnet. 
Additionally, the interfacial spin accumulation may modify the
interface
contribution to the
conductivity, which may contribute to the UMR as
well~\cite{origins_unidirectional_spin_hall_magnetoresistance}.
A thickness-dependent study~\cite{thickness_dependence_unidirectional_SHM}
in magnetic bilayers confirms the role of the SHE for the UMR.
Another indication that the SHE is very often at the heart of the UMR in metallic
bilayers comes from the observation that the UMR
correlates with the antidamping spin-orbit torque, 
but not with the field-like one~\cite{usmr_bilayers_avci}.
This picture changes if Rashba interface states dominate the
interfacial magnetotransport properties: In Fe/Ge(111) a large UMR
has been found that has been attributed to the Rashba effect of the
interface states~\cite{PhysRevB.103.064411}.
Also in heterostructures composed of a topological insulator
on a ferromagnet the Rashba-Edelstein
effect has been found to contribute to the UMR as well~\cite{unidirectional_spin_hall_and_rashba_edelstein}.

Moreover, the spin current injected into the ferromagnet due to the
SHE of the heavy metal layer may excite magnons in the 
ferromagnet. These magnons may modify the resistivity of the
ferromagnetic layer similarly to the spin-disorder contribution to
the
resistivity and thus contribute 
to the UMR~\cite{origins_unidirectional_spin_hall_magnetoresistance}.
The large UMR in topological insulator heterostructures has been
attributed to asymmetric electron-magnon
scattering~\cite{large_unidirectional_mr_magnetic_ti}. 
Finally, UMR exists not only in magnetic heterostructures but also in
bulk ferromagnets with broken inversion 
symmetry~\cite{electrical_detection_magnetization_reversal,2021unidirectionalNiMnSb},
which might require different models to describe the UMR than the heterostructures.

In addition to the UMR the nonlinear response to the applied
electric current contains also the 
nonlinear Hall effect (NLHE)~\cite{quantum_nonlin_hall_berry_curvature_dipole}.
Some mechanisms of the UMR discussed above may also lead to NLHE.
For example, 
asymmetric electron-magnon scattering has been found to contribute
to the NLHE~\cite{current_nonlinear_hall_effect}.

So far, most theoretical models of UMR address only one particular 
mechanism. 
Ref.~\cite{unidirectional_theory} develops a model to describe the UMR from the 
modulation of the conductivity of the
ferromagnetic layer when a spin current from a heavy metal layer is
injected
due to the SHE.
For magnetic bilayer systems composed of an insulating ferromagnet on
a
heavy metal layer a theoretical model was developed to describe the
magnonic contribution to the UMR~\cite{PhysRevB.99.064438}.
Ref.~\cite{2021unidirectionalNiMnSb} uses the
Boltzmann transport theory to derive an expression
for UMR, which is applied to NiMnSb.
Ref.~\cite{PhysRevResearch.2.043081} expresses the nonlinear
conductivity in the clean limit in terms of the Berry curvature dipole and a Drude term
and
applies this theory to a model of 
BaMn$_{2}$As$_{2}$. 

While UMR and NLHE are relatively new effects in spintronics, there are
much older second order responses well-known in 
nonlinear optics, e.g. the shift current, the injection current and
the 2nd harmonic generation~\cite{PhysRevB.61.5337}.
The nonlinear conductivities at optical frequencies contain also the
photovoltaic anomalous Hall effect, which has been considered 
recently in line-node semimetals~\cite{PhysRevB.94.155206},
and it also contains the
photovoltaic chiral magnetic effect, which has been studied in Weyl
semimetals
recently~\cite{PhysRevB.93.201202}.
At first glance it is tempting to guess that formulae suitable to
compute
the UMR and the NLHE may be obtained easily by taking the 
zero-frequency limit of the dc photocurrent expressions. However,
as we will discuss in this work this is not the case.
Nevertheless, it is instructive to compare the second order response
tensors derived in nonlinear optics to the expressions for UMR and NLHE.
Since second order response coefficients are considerably more
complicated
to compute than the linear ones a large number of nonlinear optics
works are devoted to the topic of comparing various formalisms and
finding
the most efficient approach for 
calculations~\cite{PhysRevB.61.5337,PhysRevB.96.035431,PhysRevB.97.205432,PhysRevB.97.235446,PhysRevB.99.045121,Jo_o_2019}.
An important conclusion of these works is that all formalisms yield
the
same answer if all caveats are considered properly.

In view of the large number of the proposed mechanisms of UMR and NLHE
it is desirable to derive general expressions for the nonlinear response
coefficients that quantify these effects. Ideally, these expressions
should
cover all possible mechanisms and they should be in a form that allows
us
to apply them within first-principles density-functional theory
calculations.
In this work we derive formulae for the second order response of
the
electric current to an applied electric field using two different
approaches:
The Keldysh nonequilibrium formalism on the one hand and the
Moyal-Keldysh formalism on the other hand. We show that these
two different formalisms lead to identical numerical results for the UMR and the
NLHE in the
ferromagnetic
Rashba model, which corroborates the applicability of both methods
to magnetic Hamiltonians with spin-orbit interaction (SOI).
In our numerical study of the Rashba model we use the independent
particle
approximation and describe effects of disorder effectively through a
quasiparticle broadening parameter. However, in our general presentation
of the Moyal-Keldysh formalism we give explicit expressions for the
self-energies,
which may be used to go beyond this constant broadening model. 
In our discussion of the UMR and NLHE in the 
ferromagnetic Rashba model we investigate the dependence on the SOI
strength,
on the Fermi energy, and on the quasiparticle broadening.
Additionally, we show analytically that the Keldysh approach converges
to the same result as an expression in the literature that was
obtained from the Boltzmann formalism
within the constant relaxation time approximation.

This paper is structured as follows:
In Sec.~\ref{sec_finite_frequency_keldysh_approach}
we use the Keldysh formalism to
derive the response coefficient for the second order in the applied electric field.
In Sec.~\ref{sec_constant_emf_keldysh_approach}
we use the Moyal-Keldysh technique to derive this response, where we
defer detailed definitions of Green functions and self energies to the Appendix~\ref{sec_appendix}.
In Sec.~\ref{sec_rashba_model} we introduce the ferromagnetic Rashba
model, which we use for the numerical study of UMR and NLHE.
In Sec.~\ref{sec_symmetry} we discuss the symmetry properties of the
UMR and the NLHE in the ferromagnetic Rashba model.
In Sec.~\ref{sec_results} we discuss the numerical results on the UMR
and the NLHE that we obtain in the ferromagnetic Rashba model using
our Keldysh and Moyal-Keldysh approaches.
This paper ends with a summary in Sec.~\ref{sec_summary}.

\section{Formalism}

\subsection{Keldysh formalism}
\label{sec_finite_frequency_keldysh_approach}
We describe the action of the applied electric field through
the time-dependent perturbation 
\bege
\delta H(t)=e\vn{v}\cdot\vn{A}(t)
\ee
 to the Hamiltonian $H$,
where $e$ is the elementary positive charge, $\vn{v}$ is the velocity operator
and 
\bege
\vn{A}(t)=
\frac{1}{2}
\left[
\frac{\vn{E}_{0} e^{-i\omega t}}{i\omega}
-
\frac{\vn{E}_{0} e^{i\omega t}}{i\omega}
\right]=
-
\frac{\vn{E}_{0} \sin(\omega t) }{\omega}
\ee
is the vector potential with 
the corresponding electric field 
\bege
\vn{E}(t)=-\frac{\partial \vn{A}(t)}{\partial t}=
\frac{\vn{E}_{0}}{2}
\left[
e^{i\omega t}
+
e^{-i\omega t}
\right]
=\vn{E}_{0} \cos(\omega t).
\ee
In the course of the following derivations we will take the limit frequency
$\omega\rightarrow 0$ below in order to extract the dc response.

The electric current density is given by
\bege
\vn{j}(t)=-\frac{e}{iV}{\rm Tr}
\left[
\vn{v} G^{<} (t,t)
\right],
\ee
where $G^{<}$ is the lesser Green function and $V$ is the volume of
the
system.
One may expand $G^{<}$ in orders of the perturbation $\delta H(t)$.
The contribution to $G^{<}$ that is quadratic in $\delta H(t)$
is given by~\cite{lasintor}
\bege\label{eq_dyson}
\begin{aligned}
&G^{<}_{2}(t,t')=\\
&\int\!\!\rmd t_1
\!\!\int\!\!\rmd t_2\,
G^{\rm R}_{\rm 0}(t,t_1)
\frac{\delta H(t_1)}{\hbar}
G^{\rm R}_{\rm 0}(t_1,t_2)
\frac{\delta H(t_2)}{\hbar}
G^{<}_{\rm 0}(t_2,t')+\\
&
\int\!\!\rmd t_1
\!\!\int\!\!\rmd t_2\,
G^{\rm R}_{\rm 0}(t,t_1)
\frac{\delta H(t_1)}{\hbar}
G^{<}_{\rm 0}(t_1,t_2)
\frac{\delta H(t_2)}{\hbar}
G_{\rm 0}^{\rm A}(t_2,t')+\\
&
\int\!\!\rmd t_1
\!\!\int\!\!\rmd t_2\,
G_{\rm 0}^{<}(t,t_1)
\frac{\delta H(t_1)}{\hbar}
G_{\rm 0}^{\rm A}(t_1,t_2)
\frac{\delta H(t_2)}{\hbar}
G_{\rm 0}^{\rm A}(t_2,t'),
\end{aligned}
\ee
where 
\bege
G^{\rm R}_{\rm 0}(t,t')=\frac{1}{2\pi\hbar}
\int_{-\infty}^{\infty}
\!\!\rmd \mathcal{E}\, \rme^{-i\mathcal{E}(t-t')/\hbar}G^{\rm R}_{\rm 0}(\mathcal{E})
\ee
is the retarded Green function in equilibrium
with Fourier 
transform $G^{\rm R}_{\rm 0}(\mathcal{E})=\hbar/[\mathcal{E}-H+i\Gamma]$.
Similarly, $G^{\rm A}_{\rm 0}(t,t_1)$ 
and $G^{<}_{\rm 0}(t,t_1)$ 
are the advanced and lesser
Green functions in equilibrium, respectively, with Fourier transforms
$G^{\rm A}_{\rm 0}(\mathcal{E})=[G^{\rm R}_{\rm 0}(\mathcal{E})]^{\dagger}$
and $G^{<}_{\rm 0}(\mathcal{E})=[G^{\rm A}_{\rm 0}(\mathcal{E})-
G^{\rm R}_{\rm 0}(\mathcal{E}) )]f(\mathcal{E})$, where $f(\mathcal{E})$
is the Fermi-Dirac distribution function. In this section we
use the independent particle approximation and assume that
lifetime effects and effects of impurity scattering can be described
by the quasiparticle broadening $\Gamma>0$. In section~\ref{sec_constant_emf_keldysh_approach}
we will give explicit expressions to compute the self-energy within
the Moyal-Keldysh approach. 

In order to evaluate the time-integrations in Eq.~\eqref{eq_dyson} 
we use
\bege
\begin{aligned}
&\int\!\!\!
 \rmd t_1\!\!\!
\int\!\!\!
\rmd t_2
G_{\rm 0}^{\eta}(t,t_1)
e^{-i\omega_1 t_1}
G_{\rm 0}^{\eta'}(t_1,t_2)
e^{-i\omega_2 t_2}
G_{\rm 0}^{\eta''}(t_2,t)
=\\
&=
\frac{\rme^{-i[\omega_1+\omega_2]t}}{h}
\!\int\!\!
\rmd \mathcal{E}
G_{\rm 0}^{\eta}(\mathcal{E}\!+\!\hbar\omega_1)
G_{\rm 0}^{\eta'}(\mathcal{E})
G_{\rm 0}^{\eta''}\!(\mathcal{E}\!-\!\hbar\omega_2),\\
\end{aligned}
\ee
where $\eta,\eta',\eta''={\rm R,A,<}$ and $\omega_1,\omega_2=\pm\omega$.
When we set
$\omega_1=\omega_2=\pm\omega$ we obtain the
$2\omega$ and $-2\omega$ contributions, while we
access the dc component by 
setting $\omega_1=-\omega_2=\pm\omega$.
Thus, the
$-2\omega$ component of the lesser Green function is given by

\bege\label{eq_lesser_min2omega}
\begin{aligned}
&G^{<}_{-2\omega}(t,t)=-
\frac{e^2}{4\omega^2 h\hbar^2}\mathcal{F}(\omega,-\omega)
e^{-2i\omega t},\\
\end{aligned}
\ee
where we defined 
\bege\label{eq_function_f}
\begin{aligned}
&\mathcal{F}(\Omega_1,\Omega_2)=\int \rmd \mathcal{E}
\\
&
\gret_{0}(\mathcal{E}+\hbar\Omega_1)
\vn{v}\cdot\vn{E}_{0}
\gret_{0}(\mathcal{E})
\vn{v}\cdot\vn{E}_{0}
\gles_{0}(\mathcal{E}+\hbar\Omega_2)+\\
+&
\gret_{0}(\mathcal{E}+\hbar\Omega_1)
\vn{v}\cdot\vn{E}_{0}
\gles_{0}(\mathcal{E})
\vn{v}\cdot\vn{E}_{0}
\gadv_{0}(\mathcal{E}+\hbar\Omega_2)+\\
+&\gles_{0}(\mathcal{E}+\hbar\Omega_1)
\vn{v}\cdot\vn{E}_{0}
\gadv_{0}(\mathcal{E})
\vn{v}\cdot\vn{E}_{0}
\gadv_{0}(\mathcal{E}+\hbar\Omega_2).\\
\end{aligned}
\ee
Similarly,
the
$2\omega$ component is given by
\bege\label{eq_lesser_2omega}
\begin{aligned}
&G^{<}_{2\omega}(t,t)=-
\frac{e^2}{4\omega^2 h\hbar^2}
\mathcal{F}(-\omega,\omega)
e^{2i\omega t},\\
\end{aligned}
\ee
and the
dc component is as follows:
\bege\label{eq_lesser_dc_noabrev}
\begin{aligned}
&G^{<}_{\rm dc}(t,t)=
\frac{e^2}{4\omega^2 h\hbar^2}
\left[
\mathcal{F}(-\omega,-\omega)
+
\mathcal{F}(\omega,\omega)
\right].
\end{aligned}
\ee

Obviously, these three components may be written as products of the $\omega^{-2}$
factor, the complex exponential, and the remainder of the expression:
$G^{<}_{-2\omega}(t,t)=\omega^{-2} g^{<}_{-2\omega}(\omega)e^{-2i\omega t}$,
$G^{<}_{2\omega}(t,t)=\omega^{-2} g^{<}_{2\omega}(\omega) e^{2i\omega t}$,
and $G^{<}_{\rm dc}(t,t)=\omega^{-2} g^{<}_{\rm dc}(\omega)$.
The sum of these three contributions may thus be formulated as
\bege\label{eq_split_timefactors}
\begin{aligned}
&[G^{<}_{-2\omega}(t,t)+G^{<}_{2\omega}(t,t)+G^{<}_{\rm dc}(t,t)]=\\
\frac{1}{\omega^2}&[g^{<}_{-2\omega}(\omega)+g^{<}_{2\omega}(\omega)+g^{<}_{\rm
  dc}(\omega)]\cos^2(\omega t)+\\
+\frac{1}{\omega^2}&[-g^{<}_{-2\omega}(\omega)-g^{<}_{2\omega}(\omega)+g^{<}_{\rm
  dc}(\omega)]\sin^2(\omega t)+\\
+\frac{i}{\omega^2}&[-g^{<}_{-2\omega}(\omega)+g^{<}_{2\omega}(\omega)]\sin(2\omega t).\\
\end{aligned}
\ee 
Next, we need to take the $\omega\rightarrow 0$ limit in order to
extract the dc response. Since the dc response is time-independent by
definition, we have the freedom to set $t$ in Eq.~\eqref{eq_split_timefactors}
to a value that makes the evaluation particularly convenient.
Therefore, we choose $t=0$, because then only the first term on the
right-hand side of Eq.~\eqref{eq_split_timefactors}
needs to be computed, because the second and third terms are zero for
$t=0$:
\bege\label{eq_split_timefactors_zerotime}
\begin{aligned}
&[G^{<}_{-2\omega}(0,0)+G^{<}_{2\omega}(0,0)+G^{<}_{\rm dc}(0,0)]=\\
\frac{1}{\omega^2}&[g^{<}_{-2\omega}(\omega)+g^{<}_{2\omega}(\omega)+g^{<}_{\rm
  dc}(\omega)].
\end{aligned}
\ee 
In order to compute the zero-frequency limit of Eq.~\eqref{eq_split_timefactors_zerotime}
we first observe 
that $\lim_{\omega\rightarrow
  0}[g^{<}_{-2\omega}(\omega)+g^{<}_{2\omega}(\omega)+g^{<}_{\rm
  dc}(\omega)]=0$,
because
\bege
\begin{aligned}
&\lim_{\omega\rightarrow 0}
\left[
-\mathcal{F}(\omega,-\omega)
-\mathcal{F}(-\omega,\omega)
+\mathcal{F}(-\omega,-\omega)
+\mathcal{F}(\omega,\omega)
\right]\\
&=
\left[
-\mathcal{F}(0,0)
-\mathcal{F}(0,0)
+\mathcal{F}(0,0)
+\mathcal{F}(0,0)
\right]=0.
\end{aligned}
\ee
Moreover, we can show that 
\bege
\begin{aligned}
&\lim_{\omega\rightarrow 0}
\frac{1}{\omega}[g^{<}_{-2\omega}(\omega)+g^{<}_{2\omega}(\omega)+g^{<}_{\rm
  dc}(\omega)]=\\
=&\lim_{\omega\rightarrow 0}
\frac{\partial}{\partial \omega}
[g^{<}_{-2\omega}(\omega)+g^{<}_{2\omega}(\omega)+g^{<}_{\rm
  dc}(\omega)]=0.
\end{aligned}
\ee
Consequently, we may use
\bege
\begin{aligned}
&\lim_{\omega\rightarrow 0}
\frac{1}{\omega^2}[g^{<}_{-2\omega}(\omega)+g^{<}_{2\omega}(\omega)+g^{<}_{\rm
  dc}(\omega)]=\\
&\lim_{\omega\rightarrow 0}
\frac{1}{2}\frac{\partial^2}{\partial \omega^2}[g^{<}_{-2\omega}(\omega)+g^{<}_{2\omega}(\omega)+g^{<}_{\rm
  dc}(\omega)].\\
\end{aligned}
\ee

These zero-frequency limits of the second derivatives
are given by
\bege\label{eq_lesser_min2omega_zerofreq}
\lim_{\omega\rightarrow 0}
\frac{1}{2}\frac{\partial^2}{\partial \omega^2}
g^{<}_{2\omega}(\omega)
=
\lim_{\omega\rightarrow 0}
\frac{1}{2}\frac{\partial^2}{\partial \omega^2}
g^{<}_{-2\omega}(\omega)
=-\Xi(-2)
\ee
and
\bege\label{eq_lesser_dc_zerofreq}
\lim_{\omega\rightarrow 0}
\frac{1}{2}\frac{\partial^2}{\partial \omega^2}
g^{<}_{\rm dc}(\omega)
=2\, \Xi(2),
\ee
where we defined the function
\bege\label{eq_define_xi_of_xi}
\begin{aligned}
&\Xi(\xi)
=
\frac{e^2}{8 h}\int \rmd \mathcal{E}\Big\{\\
&
\frac{\partial^2}{\partial \mathcal{E}^2}
\gret_{0}(\mathcal{E})
\vn{v}\cdot\vn{E}_{0}
\gret_{0}(\mathcal{E})
\vn{v}\cdot\vn{E}_{0}
\gles_{0}(\mathcal{E})+\\
&
\gret_{0}(\mathcal{E})
\vn{v}\cdot\vn{E}_{0}
\gret_{0}(\mathcal{E})
\vn{v}\cdot\vn{E}_{0}
\frac{\partial^2}{\partial \mathcal{E}^2}
\gles_{0}(\mathcal{E})+\\
&\xi
\frac{\partial}{\partial \mathcal{E}}
\gret_{0}(\mathcal{E})
\vn{v}\cdot\vn{E}_{0}
\gret_{0}(\mathcal{E})
\vn{v}\cdot\vn{E}_{0}
\frac{\partial}{\partial \mathcal{E}}
\gles_{0}(\mathcal{E})+\\
+&
\frac{\partial^2}{\partial \mathcal{E}^2}
\gret_{0}(\mathcal{E})
\vn{v}\cdot\vn{E}_{0}
\gles_{0}(\mathcal{E})
\vn{v}\cdot\vn{E}_{0}
\gadv_{0}(\mathcal{E})+\\
+&
\gret_{0}(\mathcal{E})
\vn{v}\cdot\vn{E}_{0}
\gles_{0}(\mathcal{E})
\vn{v}\cdot\vn{E}_{0}
\frac{\partial^2}{\partial \mathcal{E}^2}
\gadv_{0}(\mathcal{E})+\\
&\xi
\frac{\partial}{\partial \mathcal{E}}
\gret_{0}(\mathcal{E})
\vn{v}\cdot\vn{E}_{0}
\gles_{0}(\mathcal{E})
\vn{v}\cdot\vn{E}_{0}
\frac{\partial}{\partial \mathcal{E}}
\gadv_{0}(\mathcal{E})+\\
+&\frac{\partial^2}{\partial \mathcal{E}^2}G_{0}^{<}(\mathcal{E})
\vn{v}\cdot\vn{E}_{0}
\gadv_{0}(\mathcal{E})
\vn{v}\cdot\vn{E}_{0}
\gadv_{0}(\mathcal{E})+\\
+&G_{0}^{<}(\mathcal{E})
\vn{v}\cdot\vn{E}_{0}
\gadv_{0}(\mathcal{E})
\vn{v}\cdot\vn{E}_{0}
\frac{\partial^2}{\partial \mathcal{E}^2}
\gadv_{0}(\mathcal{E})+\\
+&\xi
\frac{\partial}{\partial \mathcal{E}}
G_{0}^{<}(\mathcal{E})
\vn{v}\cdot\vn{E}_{0}
\gadv_{0}(\mathcal{E})
\vn{v}\cdot\vn{E}_{0}
\frac{\partial}{\partial \mathcal{E}}
\gadv_{0}(\mathcal{E})
\Big\}.\\
\end{aligned}
\ee

Summing up terms, we obtain
\bege\label{eq_lesser_dc_cos_squared}
\begin{aligned}
&\lim_{\omega\rightarrow 0}
\frac{
g^{<}_{-2\omega}(\omega)+g^{<}_{2\omega}(\omega)+g^{<}_{\rm
  dc}(\omega)
}{\omega^2}
=\frac{e^2}{2 h}\int \rmd \mathcal{E}\Big\{\\
&2
\frac{\partial}{\partial \mathcal{E}}
\gret_{0}(\mathcal{E})
\vn{v}\cdot\vn{E}_{0}
\gret_{0}(\mathcal{E})
\vn{v}\cdot\vn{E}_{0}
\frac{\partial}{\partial \mathcal{E}}
\gles_{0}(\mathcal{E})+\\
&2
\frac{\partial}{\partial \mathcal{E}}
\gret_{0}(\mathcal{E})
\vn{v}\cdot\vn{E}_{0}
\gles_{0}(\mathcal{E})
\vn{v}\cdot\vn{E}_{0}
\frac{\partial}{\partial \mathcal{E}}
\gadv_{0}(\mathcal{E})+\\
+&2
\frac{\partial}{\partial \mathcal{E}}
G_{0}^{<}(\mathcal{E})
\vn{v}\cdot\vn{E}_{0}
\gadv_{0}(\mathcal{E})
\vn{v}\cdot\vn{E}_{0}
\frac{\partial}{\partial \mathcal{E}}
\gadv_{0}(\mathcal{E})
\Big\}.\\
\end{aligned}
\ee
The energy derivative of the
lesser Green function contains one
term proportional to the Fermi function
and a second term proportional to the
energy derivative of the Fermi function:
\bege
\begin{aligned}
\frac{\partial G^{<}_{\rm 0}(\mathcal{E})}{\partial \mathcal{E}}&=
\left[
\frac{\partial G^{\rm A}_{\rm 0}(\mathcal{E})}{\partial \mathcal{E}}-
\frac{\partial G^{\rm R}_{\rm 0}(\mathcal{E})}{\partial \mathcal{E}}
\right]
f(\mathcal{E})\\
&+\left[
G^{\rm A}_{\rm 0}(\mathcal{E})-
G^{\rm R}_{\rm 0}(\mathcal{E})
\right]
f'(\mathcal{E}).\\
\end{aligned}
\ee
First, we separate these contributions proportional to $f$ and $f'$ in
Eq.~\eqref{eq_lesser_dc_cos_squared}.
The terms proportional to $f$ yield
\bege\label{eq_lesser_zerofreq_proptof}
\begin{aligned}
&
\frac{e^2}{2 h}\int f(\mathcal{E}) \rmd \mathcal{E}\Big\{\\
&2
\frac{\partial}{\partial \mathcal{E}}
\gret_{0}(\mathcal{E})
\vn{v}\cdot\vn{E}_{0}
\gret_{0}(\mathcal{E})
\vn{v}\cdot\vn{E}_{0}
\frac{\partial}{\partial \mathcal{E}}
\glestwo_{0}(\mathcal{E})+\\
+&2
\frac{\partial}{\partial \mathcal{E}}
\gret_{0}(\mathcal{E})
\vn{v}\cdot\vn{E}_{0}
\glestwo_{0}(\mathcal{E})
\vn{v}\cdot\vn{E}_{0}
\frac{\partial}{\partial \mathcal{E}}
\gadv_{0}(\mathcal{E})+\\
+&2
\frac{\partial}{\partial \mathcal{E}}
\glestwo_{0}(\mathcal{E})
\vn{v}\cdot\vn{E}_{0}
\gadv_{0}(\mathcal{E})
\vn{v}\cdot\vn{E}_{0}
\frac{\partial}{\partial \mathcal{E}}
\gadv_{0}(\mathcal{E})
\Big\},\\
&
=\frac{e^2}{ h}\int f(\mathcal{E}) \rmd \mathcal{E}\Big\{\\
&
\frac{\partial
\gadv_{0}(\mathcal{E})
}{\partial \mathcal{E} }
\vn{v}\cdot\vn{E}_{0}
\gadv_{0}(\mathcal{E})
\vn{v}\cdot\vn{E}_{0}
\frac{\partial
\gadv_{0}(\mathcal{E})
}{\partial \mathcal{E} }
\\
-&
\frac{\partial
\gret_{0}(\mathcal{E})
}{\partial \mathcal{E} }
\vn{v}\cdot\vn{E}_{0}
\gret_{0}(\mathcal{E})
\vn{v}\cdot\vn{E}_{0}
\frac{\partial
\gret_{0}(\mathcal{E})
}{\partial \mathcal{E} }
\Big\},\\
\end{aligned}
\ee
where we defined
\bege
\glestwo_{0}(\mathcal{E})=\gadv_{0}(\mathcal{E})-\gret_{0}(\mathcal{E}).
\ee
We introduce the second-order conductivity tensor $\sigma_{\alpha\beta\gamma}$ through
\bege
J_{\alpha}=\sum_{\beta\gamma}\sigma_{\alpha\beta\gamma}E_{\beta}E_{\gamma}.
\ee
From Eq.~\eqref{eq_lesser_zerofreq_proptof} we
obtain the following contribution to $\sigma_{\alpha\beta\gamma}$:
\bege\label{eq_sigma_I}
\begin{aligned}
&\sigma^{\rm (I)}_{\alpha\beta\gamma}=
\frac{2e^3}{hV}
\int f(\mathcal{E}) \rmd \mathcal{E}
{\rm Im}
\Big\{
{\rm Tr}
[
\\
&v_{\alpha}
\frac{\partial
\gret_{0}(\mathcal{E})
}{\partial \mathcal{E} }
v_{\beta}
\gret_{0}(\mathcal{E})
v_{\gamma}
\frac{\partial
\gret_{0}(\mathcal{E})
}{\partial \mathcal{E} }
] \Big\}.
\\
\end{aligned}
\ee

Additionally, 
the terms proportional to $f'$ produce the contribution
\bege\label{eq_lesser_zerofreq_fprime}
\begin{aligned}
&
\frac{e^2}{2 h}
\int 
f'(\mathcal{E})
\rmd \mathcal{E}\Big\{\\
&
2\frac{\partial}{\partial \mathcal{E}}
\gret_{0}(\mathcal{E})
\vn{v}\cdot\vn{E}_{0}
\gret_{0}(\mathcal{E})
\vn{v}\cdot\vn{E}_{0}
\glestwo_{0}(\mathcal{E})+\\
&
2
\glestwo_{0}(\mathcal{E})
\vn{v}\cdot\vn{E}_{0}
\gadv_{0}(\mathcal{E})
\vn{v}\cdot\vn{E}_{0}
\frac{\partial}{\partial \mathcal{E}}
\gadv_{0}(\mathcal{E})\Big\}\\
\end{aligned}
\ee
to the lesser Green function, which contributes
\bege\label{eq_sigma_II}
\begin{aligned}
&\sigma^{\rm (II)}_{\alpha\beta\gamma}=-\frac{2e^3}{h V}
\int 
f'(\mathcal{E}) \rmd \mathcal{E}
{\rm Im}\Big\{{\rm Tr}
[\\
&
v_{\alpha}
\frac{\partial \gret_{0}(\mathcal{E})}{\partial \mathcal{E}}
v_{\beta}
\gret_{0}(\mathcal{E})
v_{\gamma}
\glestwo_{0}(\mathcal{E})
]
\Big\}
\end{aligned}
\ee
to the conductivity.
The total second-order conductivity tensor is given 
by
\bege\label{eq_keldysh_total}
\sigma^{\phantom{I}}_{\alpha\beta\gamma}=\sigma^{\rm (I)}_{\alpha\beta\gamma}+\sigma^{\rm (II)}_{\alpha\beta\gamma}.
\ee

When we compare the derivations in this section to the
derivations of the expressions for the 
laser-induced dc torque~\cite{lasintor} 
and dc photocurrent~\cite{lasincucspira}
within the Keldysh nonequilibrium formalism
we observe that $\sigma_{\alpha\beta\gamma}$ in Eq.~\eqref{eq_keldysh_total} is not simply related
to the $\omega\rightarrow 0$ limit of the dc photocurrent.
The reason for this is that we have to discard the 2nd harmonics in the
derivation of the dc photocurrent. However, in the derivation of 
the expression for $\sigma_{\alpha\beta\gamma}$ we cannot discard the 2nd harmonics,
because they contribute to the dc response as $\omega\rightarrow 0$.
The observation that the second-order dc response is not simply the
$\omega\rightarrow 0$ limit of the dc photocurrent
may also be obtained by a different argument:
The photocurrent and the inverse Faraday effect
may diverge as $\omega\rightarrow 0$~\cite{PhysRevX.10.041041,PhysRevLett.117.137203}, while
the
second-order response to a dc electric field has to be finite.

In order to apply Eq.~\eqref{eq_sigma_I}
and Eq.~\eqref{eq_sigma_II}
to periodic solids
we introduce periodic boundary conditions. As a consequence,
the Hamiltonian and the Green function become dependent on the
$k$-point, but we do not write this
$k$-dependence explicitly in the equations for notational convenience.
We introduce a $k$-integration by replacing the system volume
in Eq.~\eqref{eq_sigma_I}
and Eq.~\eqref{eq_sigma_II} as follows:
\bege
\label{eq_modify_for_k_integ}
\frac{1}{V}\dots \rightarrow \int \frac{\rmd^D k}{(2\pi)^D}\dots,
\ee
where $D$ is the dimension of the system.

Important contributions to Eq.~\eqref{eq_keldysh_total} are given
by the intraband terms:
\bege\label{eq_keldysh_intraband}
\begin{aligned}
&\sigma^{\rm (II,intra)}_{\alpha\beta\gamma}=-\frac{2e^3\hbar^2}{\pi }
\int \frac{\rmd^D k}{(2\pi)^D}
\sum_{n}
v_{\alpha,\vn{k}n}
v_{\beta,\vn{k}n}
v_{\gamma,\vn{k}n}\times\\
&\times\int 
f'(\mathcal{E}) \rmd \mathcal{E}
\frac{\Gamma (\mathcal{E}-\mathcal{E}_{\vn{k}n})}{[(\mathcal{E}-\mathcal{E}_{\vn{k}n})^2+\Gamma^2]^3},
\end{aligned}
\ee
where $v_{\alpha,\vn{k}n}=\langle \vn{k} n| v_{\alpha} | \vn{k}n\rangle$ are the intraband matrix
elements of the velocity operator, $\mathcal{E}_{\vn{k}n}$ is the band
energy of an electron in band $n$ at $k$-point $\vn{k}$ and $|\vn{k}n\rangle$ is the
corresponding state.
Using integration by parts we may
rewrite Eq.~\eqref{eq_keldysh_intraband}
as follows:
\bege\label{eq_keldysh_intraband_ibp}
\begin{aligned}
&\sigma^{\rm (II,intra)}_{\alpha\beta\gamma}=-\frac{e^3\hbar^2}{2\pi}
\int \frac{\rmd^D k}{(2\pi)^D}
\sum_{n}
v_{\alpha,\vn{k}n}
v_{\beta,\vn{k}n}
v_{\gamma,\vn{k}n}\times\\
&\times\int 
f''(\mathcal{E}) \rmd \mathcal{E}
\frac{\Gamma }{[(\mathcal{E}-\mathcal{E}_{\vn{k}n})^2+\Gamma^2]^2}.
\end{aligned}
\ee
In the limit $\Gamma\rightarrow 0$ this turns into
\bege\label{eq_keldysh_intraband_ibp_gzero}
\begin{aligned}
&\sigma^{\rm (II,intra)}_{\alpha\beta\gamma}=-e^3\tau^2
\int \frac{\rmd^D k}{(2\pi)^D}
\sum_{n}
v_{\alpha,\vn{k}n}
v_{\beta,\vn{k}n}
v_{\gamma,\vn{k}n}
f''(\mathcal{E}_{\vn{k}n}),
\end{aligned}
\ee
where $\tau=\hbar/(2\Gamma)$ is the relaxation time.

Interestingly, Eq.~\eqref{eq_keldysh_intraband_ibp_gzero}
differs from the Boltzmann result~\cite{2021unidirectionalNiMnSb}
\bege \label{eq_boltzmann}
\sigma^{(\rm Boltz )}_{\alpha\beta\gamma}=-\frac{e^3\tau^2}{2}
\int \frac{\rmd^D k}{(2\pi)^D}
\sum_{n}
v_{\alpha,\vn{k}n}
v_{\beta,\vn{k}n}
v_{\gamma,\vn{k}n}
f''(\mathcal{E}_{\vn{k}n})
\ee
by a factor of 2. This means that the intraband terms
in the Keldysh approach do not directly correspond to the
Boltzmann result. The reason for this is that the
f-sum rule~\cite{oms_ogata_fukuyama} 
\bege
\left[
\frac{1}{\hbar^2}\frac{\partial^2 \mathcal{E}_{\vn{k}l}}{\partial k_{\alpha}\partial
  k_{\beta}}
-\frac{\delta_{\alpha\beta}}{m^{*}}
\right]
=
2\sum_{n\ne l}
\frac{
\left\langle 
\vn{k}l 
\left|
v_{\alpha}
\right|
\vn{k}n
\right\rangle
\left\langle 
\vn{k}n
\left|
v_{\beta}
\right|
\vn{k}l
\right\rangle
}{
\mathcal{E}_{\vn{k}l}-\mathcal{E}_{\vn{k}n}
}
\ee
allows us to transform intraband terms into interband terms
and vice versa: The left-hand side of this expression
seems to be an intraband term, while the right-hand side
of this expression seems to be an interband term.
This means that the terms 'intraband' and 'interband' need
to be used with care, because the one kind may be transformed into the
other kind~\cite{oms_ogata_fukuyama}. In this expression
$m^{*}$ is the mass used in the expression of the kinetic energy in
the
Hamiltonian. Thus, in the framework of \textit{ab-initio} density-functional theory
calculations $m^{*}$ is the electron mass, i.e., $m^{*}=m_{\rm e}$.
However, in the Rashba model, which we discuss in Sec.~\ref{sec_rashba_model},
$m^{*}$ is a free parameter that can be tuned to model the band dispersion.

While the intraband terms of the rigorous
quantum mechanical first-order perturbation theory usually
correspond to the
semiclassical approach, this is not true any more for the
higher order perturbation theory. Already in the second order
perturbation theory the f-sum rule needs to be used in order to
connect the rigorous quantum mechanical approach to the 
semiclassical one in the case of the orbital magnetic
susceptibility~\cite{oms_ogata_fukuyama}.

In order to see that the f-sum rule allows us to resolve the
discrepancy between Eq.~\eqref{eq_keldysh_intraband_ibp_gzero}
and Eq.~\eqref{eq_boltzmann} we consider the
identity
\bege
\lim_{\Gamma\rightarrow 0}{\rm Im}
\int 
\frac{\Gamma^2 f'(\mathcal{E}) \rmd \mathcal{E}}{[\mathcal{E}-\mathcal{E}_{\vn{k}n}+i\Gamma]^2[\mathcal{E}-\mathcal{E}_{\vn{k}n}-i\Gamma]}
=-\frac{\pi}{2}f'(\mathcal{E}_{\vn{k}n}),
\ee
which can be used to show that Eq.~\eqref{eq_sigma_II}
contains precisely one interband term that scales like $\Gamma^{-2}$
in the clean limit, namely
\bege\label{eq_sigma_II_interband}
\begin{aligned}
\sigma^{\rm (II,inter)}_{\alpha\beta\gamma}&
=\frac{e^3\hbar^2}{\pi}
\int \frac{\rmd^D k}{(2\pi)^D}
\int 
f'(\mathcal{E}) \rmd \mathcal{E}
\sum_{n}\sum_{m\ne n}{\rm Im}\Big\{\\
&
\frac{
\langle \vn{k}n|
v_{\alpha}
| \vn{k}n \rangle
\langle \vn{k}n|
v_{\beta}
| \vn{k}m\rangle
\langle \vn{k} m|
v_{\gamma}
|\vn{k}n\rangle
}{(\mathcal{E}-\mathcal{E}_{\vn{k}n}+i\Gamma)^2
(\mathcal{E}-\mathcal{E}_{\vn{k}n}-i\Gamma)
 (\mathcal{E}-\mathcal{E}_{\vn{k}m}+i\Gamma)}
\Big\}\\
&
\simeq
-\frac{e^3\hbar^2}{2\Gamma^2 }
\int \frac{\rmd^D k}{(2\pi)^D}
\sum_{n}\sum_{m\ne n}
f'(\mathcal{E}_{\vn{k}n}) \times
\\
&\times
\langle \vn{k}n|
v_{\alpha}
| \vn{k}n \rangle
{\rm Re}
\left[
\frac{
\langle \vn{k}n|
v_{\beta}
| \vn{k}m\rangle
\langle \vn{k} m|
v_{\gamma}
|\vn{k}n\rangle
}{(\mathcal{E}_{\vn{k}n}-\mathcal{E}_{\vn{k}m})}
\right].\\
\end{aligned}
\ee
Using the f-sum rule we may rewrite $\sigma^{\rm (II,inter)}_{\alpha\beta\gamma}$ as
\bege\label{eq_sigma_II_interband_fsum}
\begin{aligned}
\sigma^{\rm (II,inter)}_{\alpha\beta\gamma}&
\simeq
-\frac{e^3}{4\Gamma^2 }
\int \frac{\rmd^D k}{(2\pi)^D}
\sum_{n}
f'(\mathcal{E}_{\vn{k}n})\times 
\\
&
\times\langle \vn{k}n|
v_{\alpha}
| \vn{k}n \rangle
\frac{\partial^2 \mathcal{E}_{\vn{k}n}}{\partial k_{\beta} \partial k_{\gamma}}.\\
\end{aligned}
\ee
Employing integration by parts we obtain first
\bege
\sigma^{\rm (II,inter)}_{\alpha\beta\gamma}=\sigma^{\rm (II,inter)}_{\gamma\beta\alpha}
\ee
and subsequently
\bege\label{eq_keldysh_interband_ibp_gzero}
\begin{aligned}
&\sigma^{\rm (II,inter)}_{\alpha\beta\gamma}=\frac{e^3\tau^2}{2}
\int \frac{\rmd^D k}{(2\pi)^D}
\sum_{n}
v_{\alpha,\vn{k}n}
v_{\beta,\vn{k}n}
v_{\gamma,\vn{k}n}
f''(\mathcal{E}_{\vn{k}n}).
\end{aligned}
\ee
It follows that
\bege\label{eq_final_boltzmann}
\sigma^{(\rm Boltz )}_{\alpha\beta\gamma}=\sigma^{\rm (II,intra)}_{\alpha\beta\gamma}+\sigma^{\rm (II,inter)}_{\alpha\beta\gamma},
\ee
which proves that the Boltzmann formalism and the Keldysh formalism
within the independent particle approximation
yield identical results in the limit $\tau\rightarrow \infty$, i.e., 
in the limit $\Gamma\rightarrow 0$.

\subsection{Moyal-Keldysh approach}
\label{sec_constant_emf_keldysh_approach}
Using the Moyal product
Ref.~\cite{theory_noneq_const_emf}
expands the Dyson equation
in static electric and magnetic fields.
A compact and general expression for the nonequilibrium Green function is given,
which describes the perturbation by static electric and magnetic
fields up to any required order in this perturbation.
In the following we evaluate this general expression for the
nonequilibrium Green function from Ref.~\cite{theory_noneq_const_emf}
at the second order in the applied electric field in order to
obtain an expression for the nonlinear response of the electric
current to the applied electric field. 

This Moyal-Keldysh approach
differs from the Keldysh approach in Sec.~\ref{sec_finite_frequency_keldysh_approach}
in two major aspects: In Sec.~\ref{sec_finite_frequency_keldysh_approach} we consider a
spatially homogeneous time-dependent electric field and take the
zero-frequency limit towards the end of the derivations. 
Therefore, we use a spatially homogeneous vector potential to describe
the
perturbation by the electric field (in works on
nonlinear
optics this choice is often referred to as 
the 'velocity gauge'~\cite{PhysRevB.96.035431,PhysRevB.97.205432,PhysRevB.97.235446,PhysRevB.99.045121,Jo_o_2019}).
In contrast,
the Moyal-Keldysh approach of Ref.~\cite{theory_noneq_const_emf}
considers the perturbation by static electromagnetic fields without
taking
any zero-frequency limit. In this approach the perturbation by a
spatially
homogeneous electric field is therefore described by a spatially
inhomogeneous
scalar potential (in works on
nonlinear
optics this choice is often referred to as 
the 'length gauge'~\cite{PhysRevB.96.035431,PhysRevB.97.205432,PhysRevB.97.235446,PhysRevB.99.045121}).
The difficulty of dealing with a spatially
inhomogeneous
non-periodic perturbation in the context of an infinite periodic crystal
 is solved elegantly in Ref.~\cite{theory_noneq_const_emf} through the
 use
of the Moyal product. Due to these two major differences, namely the
time-dependence on the one hand and the use of the Moyal product
on the other hand, the derivations in Sec.~\ref{sec_finite_frequency_keldysh_approach}
are quite distinct from the formalism described in this section.
In the results section we will show that these two rather distinct
approaches yield identical numerical results. We therefore present
both techniques in this manuscript, because they corroborate each
other
and thereby demonstrate the validity of both approaches for 
magnetic Hamiltonians with SOI.

Similar comparisons between the velocity gauge and the
length gauge have been done for nonlinear optical 
responses~\cite{PhysRevB.61.5337,PhysRevB.96.035431,PhysRevB.97.205432,PhysRevB.97.235446,PhysRevB.99.045121}). Some
of these works stress the advantages of the length gauge approach,
while others stress those of the velocity gauge
approach. Ref.~\cite{PhysRevB.99.045121}
advertises the velocity gauge as the more convenient choice in the
context of a diagrammatic approach. However, there is also a
diagrammatic approach to the Moyal technique (see Appendix C in
Ref.~\cite{phase_space_berry} for an illustration of several diagrams) and therefore
the length gauge may be implemented diagrammatically as well
if the Moyal technique is used.

Ref.~\cite{theory_noneq_const_emf}
provides the following expansion of the Green's function in orders of the
electromagnetic field tensor $F^{\mu\nu}$:
\bege\label{eq_expand_g_fij}
\ghat=
\ghat_{0}+
\frac{\hbar e}{2}\ghat_{\mu\nu}F^{\mu\nu}+
\frac{\hbar^2 e^2}{8}\ghat_{\mu\nu,\mu'\nu'}F^{\mu\nu}F^{\mu'\nu'}+\dots,
\ee
where
\bege
\ghat(\mathcal{E})=
\begin{pmatrix}
\gret(\mathcal{E}) &2\gles(\mathcal{E})\\
0 &\gadv(\mathcal{E})\\
\end{pmatrix}
\ee
and
\bege
\sigmahat=
\begin{pmatrix}
\sigmaret(\mathcal{E}) &2\sigmales(\mathcal{E})\\
0 &\sigmaadv(\mathcal{E})\\
\end{pmatrix}
\ee
are the Green function and the self energy in matrix form,
respectively.
In the first order, the electromagnetic field tensor contributes
\bege
\begin{aligned}
\ghat_{\mu\nu}=&\ghat_{0}\sigmahat_{\mu\nu}\ghat_{0}
-\frac{i}{2}
\ghat_{0}
\partial_{\pi^{\mu}}
\ghat_{0}^{-1}
\partial_{\pi^{\nu}}
\ghat_{0}+\\
&+
\frac{i}{2}
\ghat_{0}
\partial_{\pi^{\nu}}
\ghat_{0}^{-1}
\partial_{\pi^{\mu}}
\ghat_{0}
\end{aligned}
\ee
to the Green function
and in the second order it contributes
\bege
\begin{aligned}
&\ghat_{\mu\nu,\mu'\nu'}=\ghat_{0}
\sigmahat_{\mu\nu,\mu'\nu'}\ghat_{0}+\\
&+\ghat_{0}\sigmahat_{\mu\nu}\ghat_{\mu'\nu'}+
\ghat_{0}\sigmahat_{\mu'\nu'}\ghat_{\mu\nu}\\
&+\frac{i}{2}\ghat_{0}\partial_{\pi^{\mu}}
\sigmahat_{\mu'\nu'}\partial_{\pi^{\nu}}\ghat_{0}-
\frac{i}{2}
\ghat_{0}
\partial_{\pi^{\mu}}\ghat_{0}^{-1}
\partial_{\pi^{\nu}}\ghat_{\mu'\nu'}\\
&-\frac{i}{2}\ghat_{0}\partial_{\pi^{\nu}}
\sigmahat_{\mu'\nu'}\partial_{\pi^{\mu}}\ghat_{0}+
\frac{i}{2}
\ghat_{0}
\partial_{\pi^{\nu}}\ghat_{0}^{-1}
\partial_{\pi^{\mu}}\ghat_{\mu'\nu'}\\
&+\frac{i}{2}\ghat_{0}\partial_{\pi^{\mu'}}
\sigmahat_{\mu\nu}\partial_{\pi^{\nu'}}\ghat_{0}-
\frac{i}{2}
\ghat_{0}
\partial_{\pi^{\mu'}}\ghat_{0}^{-1}
\partial_{\pi^{\nu'}}\ghat_{\mu\nu}\\
&-\frac{i}{2}\ghat_{0}\partial_{\pi^{\nu'}}
\sigmahat_{\mu\nu}\partial_{\pi^{\mu'}}\ghat_{0}+
\frac{i}{2}
\ghat_{0}
\partial_{\pi^{\nu'}}\ghat_{0}^{-1}
\partial_{\pi^{\mu'}}\ghat_{\mu\nu}\\
&+\frac{1}{4}
\ghat_{0}\partial_{\pi^{\mu}}\partial_{\pi^{\mu'}}
\ghat_{0}^{-1}
\partial_{\pi^{\nu}}\partial_{\pi^{\nu'}}\ghat_{0}\\
&-\frac{1}{4}
\ghat_{0}\partial_{\pi^{\nu}}\partial_{\pi^{\mu'}}
\ghat_{0}^{-1}
\partial_{\pi^{\mu}}\partial_{\pi^{\nu'}}\ghat_{0}\\
&-\frac{1}{4}
\ghat_{0}\partial_{\pi^{\mu}}\partial_{\pi^{\nu'}}
\ghat_{0}^{-1}
\partial_{\pi^{\nu}}\partial_{\pi^{\mu'}}\ghat_{0}\\
&+\frac{1}{4}
\ghat_{0}\partial_{\pi^{\nu}}\partial_{\pi^{\nu'}}
\ghat_{0}^{-1}
\partial_{\pi^{\mu}}\partial_{\pi^{\mu'}}\ghat_{0},\\
\end{aligned}
\ee
where $\pi^{\mu}=
(\pi^{0},\pi^{1},\pi^{2},\pi^{3})=
(\mathcal{E}/c,\hbar\vn{k})$ is the 4-momentum ($c$ is the velocity of
light),
$\partial_{\pi^{\mu}}=\partial/\partial\pi^{\mu}$ is the corresponding
derivative,
and $\ghat_{0}$ is the equilibrium Green function in matrix form.
In order to obtain the Green function at the second order
in the electric field we set
$F^{i 0}=E_{i}/c$ and $F^{0i}=-E_{i}/c$, which simplifies
Eq.~\eqref{eq_expand_g_fij} to
\bege
\ghat=\ghat_{0}+
\frac{\hbar e}{c} 
\ghat_{E_i}E_{i}+
\frac{\hbar^2 e^2}{2 c^2}
\ghat_{E_i,E_j}E_iE_j+\dots,
\ee
where
\bege\label{eq_ghat_ee}
\begin{aligned}
&\ghat_{E_i,E_j}=
\ghat_{0}\Biggl[
\sigmahat_{E_i,E_j}
\ghat_{0}+
\sigmahat_{E_i}
\ghat_{E_j}+
\sigmahat_{E_j}
\ghat_{E_i}+\\
&-\frac{i}{2}
\partial_{\pi^i}\sigmahat_{E_j}
\partial_{\pi^0}\ghat_{0}
+\frac{i}{2}
\partial_{\pi^i}\ghat_{0}^{-1}
\partial_{\pi^0}\ghat_{E_j}\\
&+\frac{i}{2}
\partial_{\pi^{0}}
\sigmahat_{E_j}
\partial_{\pi^i}
\ghat_{0}
-\frac{i}{2}
\partial_{\pi^{0}}
\ghat_{0}^{-1}
\partial_{\pi^i}
\ghat_{E_j}\\
&-\frac{i}{2}
\partial_{\pi^j}
\sigmahat_{E_i}
\partial_{\pi^0}
\ghat_{0}
+\frac{i}{2}
\partial_{\pi^j}
\ghat_{0}^{-1}
\partial_{\pi^0}
\ghat_{E_i}\\
&+\frac{i}{2}
\partial_{\pi^0}
\sigmahat_{E_i}
\partial_{\pi^j}
\ghat_{0}
-\frac{i}{2}
\partial_{\pi^0}
\ghat_{0}^{-1}
\partial_{\pi^j}
\ghat_{E_i}\\
&+\frac{1}{4}
\partial_{\pi^i}\partial_{\pi^j}
\ghat_{0}^{-1}
\partial_{\pi^0}\partial_{\pi^0}
\ghat_{0}\\
&-\frac{1}{4}
\partial_{\pi^0}\partial_{\pi^j}
\ghat_{0}^{-1}
\partial_{\pi^0}\partial_{\pi^i}
\ghat_{0}\\
&-\frac{1}{4}
\partial_{\pi^0}\partial_{\pi^i}
\ghat_{0}^{-1}
\partial_{\pi^0}\partial_{\pi^j}
\ghat_{0}\\
&+\frac{1}{4}
\partial_{\pi^0}\partial_{\pi^0}
\ghat_{0}^{-1}
\partial_{\pi^j}\partial_{\pi^i}
\ghat_{0}\Biggr].\\
\end{aligned}
\ee

In Eq.~\eqref{eq_ghat_ee} up to
two energy derivatives $\partial_{\pi^0}$
may act on the Green functions. 
The second energy derivative of
$G^{<}_{\rm 0}(\mathcal{E})=[G^{\rm A}_{\rm 0}(\mathcal{E})-
G^{\rm R}_{\rm 0}(\mathcal{E}) )]f(\mathcal{E})$
generates terms proportional to $f$, to $f'$ and to
$f''$. Consequently, the
lesser-component of $\ghat_{E_i,E_j}$
may be written as
\bege
\label{eq_efiefi_lesser_all}
\gles_{E_i,E_j}=
f'(\mathcal{E})\glesone_{E_i,E_j}
+f(\mathcal{E})\glestwo_{E_i,E_j}
+f''(\mathcal{E})\glesthree_{E_i,E_j}.
\ee
According to Eq.~\eqref{eq_sigma_I}  and
Eq.~\eqref{eq_sigma_II} the Keldysh formalism
in the previous section does not yield a term
proportional to $f''$ at first. However, already in
Eq.~\eqref{eq_keldysh_intraband_ibp} we have 
shown that integration by parts
leads to terms proportional 
to $f''$. Conversely, we may use integration by
parts to rewrite the term involving $f''$
in Eq.~\eqref{eq_efiefi_lesser_all} as a term
proportional to $f'$. Therefore, the separation
into terms proportional to 
$f$, $f'$, and $f''$ is ambiguous rather than unique.
Consequently, when comparing the two formalisms numerically
in Sec.~\ref{sec_results} we only compare the total second order
conductivities rather than their separation into terms proportional to
$f$, $f'$, and $f''$.

Finally, the second order conductivity 
in the Moyal-Keldysh approach 
may be written as
\bege\label{eq_MoyKel_conductivity}
\sigma_{\alpha\beta\gamma}=-\frac{\hbar^2 e^3}{4\pi iV}
\int\rmd \mathcal{E}
{\rm Tr}
\left[
v_{i}
\gles_{E_i,E_j}(\mathcal{E})
\right],
\ee
where $\gles_{E_i,E_j}(\mathcal{E})$ is given
by Eq.~\eqref{eq_efiefi_lesser_all}.

Detailed expressions of the self energies 
$\sigmaret_{E_i,E_j}$, $\sigmaret_{E_i}$,
$\sigmalesone_{E_i,E_j}$, $\sigmalesone_{E_i}$, $\sigmalestwo_{E_i}$, $\sigmalesthree_{E_i,E_j}$
and of 
several Green functions are given in the Appendix~\ref{sec_appendix}.
In Eq.~\eqref{eq_retarded_ee},
Eq.~\eqref{eq_glesone_ee}, and
Eq.~\eqref{eq_glesthree_ee}
we have provided the general expressions for
$\gret_{E_i,E_j}(\mathcal{E})$,
$\glesone_{E_i,E_j}(\mathcal{E})$, and
$\glesthree_{E_i,E_j}(\mathcal{E})$,
respectively, which determine $\gles_{E_i,E_j}(\mathcal{E})$
according to Eq.~\eqref{eq_efiefi_lesser_all}.
However, for the numerical calculations in this manuscript we only use a
constant broadening $\Gamma$. Consequently, we set the
self-energies $\sigmaret_{E_i,E_j}$, $\sigmaret_{E_i}$,
$\sigmalesone_{E_i,E_j}$, $\sigmalesone_{E_i}$, $\sigmalestwo_{E_i}$, $\sigmalesthree_{E_i,E_j}$ 
to zero, which simplifies the 
Eq.~\eqref{eq_retarded_ee},
Eq.~\eqref{eq_glesone_ee}, and
Eq.~\eqref{eq_glesthree_ee}
significantly.

In the Keldysh approach a major part of the derivations 
is devoted to evaluating the limit $\omega\rightarrow 0$  
as Sec.~\ref{sec_finite_frequency_keldysh_approach} shows.
We suspect that with increasing order of the perturbation by the
electric field
taking this dc limit will become more and more cumbersome.
In contrast, in the Moyal-Keldysh approach used in this section the zero-frequency dc response
is obtained directly.
This is a major advantage of the Moyal-Keldysh approach over
the standard Keldysh approach in applications to the zero-frequency dc
response. 

\subsection{Rashba model}
\label{sec_rashba_model}
In this work we compute UMR and NLHE in the magnetic 
Rashba model~\cite{rashba_review}
\bege\label{eq_rashba_model}
H_{\vn{k}}=\frac{\hbar^2}{2m^{*}}k^2+
\alpha^{\rm R} (\vn{k}\times\hat{\vn{e}}_{z})\cdot\vn{\sigma}+
\frac{\Delta V}{2}\vn{\sigma}\cdot\hat{\vn{M}},
\ee
where $\alpha^{\rm R}$ is the Rashba parameter,
$\hat{\vn{M}}$ is the magnetization direction,
and $\Delta V$
is the exchange splitting. 
The mass $m^{*}$ may be tuned to match the band dispersion of a given
interfacial or surface state.
The electrons are constrained to move in the
$xy$ plane, i.e., $\vn{k}=(k_x,k_y,0)^{\rm T}$ and $z=0$.
The Rashba model is suitable to describe the UMR from interfacial Rashba
states~\cite{PhysRevB.103.064411}.
The effects of injection of spin-current generated in one
region into a second region are not captured by the
Rashba model, because it describes only a
single
homogeneous two-dimensional region.

When the magnetization points in the $x$ direction,
i.e., $\hat{\vn{M}}=\hat{\vn{e}}_{x}$,
the eigenenergies of $H_{\vn{k}}$ at $k$-points
$\vn{k}=(0,k_y,0)^{\rm T}$ 
and $-\vn{k}=(0,-k_y,0)^{\rm T}$ 
differ.
This $\vn{k}$ vs $-\vn{k}$ asymmetry has been 
observed in angle-resolved photoemission spectroscopy 
experiments~\cite{PhysRevB.93.125409} and it has been
suggested that it influences electron transport 
properties, namely there should be a difference in electron
transport depending on whether the current is 
applied in the $y$ direction or in the $-y$ direction. Explicitly,
Ref.~\cite{PhysRevB.93.125409} suggests that the applied
current leads to a torque on the magnetization that changes
the resisitivity due to the anisotropic magnetoresistance. As a
consequence, a voltage component quadratic in the applied
electric current is predicted, which indeed means that the
resistivity depends on whether the current is applied in the $y$
direction 
or in the $-y$ direction.

Additionally, the resistivity 
is expected to depend on whether the current 
is applied in the $y$
direction 
or in the $-y$ direction also due to the UMR. 
When measuring the UMR one therefore needs to make sure
that the magnetization direction is fixed in order to avoid the
contribution from the modulation of the magnetoresistance by 
the current-induced torque described in
Ref.~\cite{PhysRevB.93.125409}.
Nevertheless, the UMR in the Rashba model is still related
to the $\vn{k}$ vs $-\vn{k}$ asymmetry.
Similarly, the nonlinear transverse response
of the electric current may contain contributions from two different
kinds of effects: The current-induced spin-orbit torque may
modulate the anomalous Hall effect and additionally
there may be an NLHE~\cite{current_nonlinear_hall_effect}.

In order to apply Eq.~\eqref{eq_sigma_I}, Eq.~\eqref{eq_sigma_II},
and Eq.~\eqref{eq_MoyKel_conductivity} to the Rashba model we
 introduce a $\vn{k}$ integration 
according to Eq.~\eqref{eq_modify_for_k_integ}
with $D=2$.

\subsection{Symmetry}
\label{sec_symmetry}
In the following we discuss the constraints on the UMR and NLHE 
currents in the Rashba model imposed by symmetry.
We consider an electric current induced at the second order of an
applied electric field.
In the non-magnetic case, i.e., when
$\Delta V=0$ in Eq.~\eqref{eq_rashba_model}, symmetry forbids
an electric current quadratic in the applied electric field: 
A $c_2$ rotation around the $z$ direction
inverts the induced electric current, consequently it has to vanish.
For the same reason there is no quadratic response of the electric
current
in the
magnetic case ($\Delta V \ne 0$) when the 
magnetization is out-of-plane, i.e., along the $z$ direction. 

Next, we consider the
magnetic case with magnetization in-plane in the $x$ direction. 
When the electric
field is applied in the $x$ direction, or in the $y$ direction, 
no $J_{x}$ is expected, because the
$yz$ mirror plane flips the response-current but not the magnetization.
However, the $zx$ mirror plane does not forbid $J_y$ if it is 
odd in $\hat{\vn{M}}$, i.e., $\sigma_{211}$ and $\sigma_{222}$ are 
allowed by symmetry. 
Since $\sigma_{211}$ describes a response current transverse to the
applied electric field, we call it an NLHE. In contrast, the component
$\sigma_{222}$
describes a UMR. 
For the analysis of experiments, UMR is defined as a resistivity that
changes sign when the direction of the electric current is reversed
and also when the magnetization direction is reversed~\cite{unidirectiona_Avci}.
Our description of UMR by a second order response coefficient
$\sigma_{\alpha\beta\gamma}$
automatically satisfies the first requirement in this definition.
The second requirement, namely the sign change when the magnetization
direction is reversed, is met by the coefficient $\sigma_{222}$ in the
Rashba model due to symmetry:
$\sigma_{222}(\hat{\vn{M}})=-\sigma_{222}(-\hat{\vn{M}})$
because the $zx$ mirror plane forbids contributions to $\sigma_{222}(\hat{\vn{M}})$
that are even in $\hat{\vn{M}}$.

When the electric field is applied in the direction of
$[\hat{e}_{x}+\hat{e}_{y}]/\sqrt{2}$,
the $zx$ mirror plane 
modifies the electric field direction into
$[\hat{e}_{x}-\hat{e}_{y}]/\sqrt{2}$,
it flips the magnetization, while it preserves  $J_x$.
Thus, $\sigma_{112}$ and $\sigma_{121}$
are allowed  by symmetry,
if they are odd in $\hat{\vn{M}}$.
The $yz$ mirror plane 
modifies the $[\hat{e}_{x}+\hat{e}_{y}]/\sqrt{2}$ direction 
of the electric field into
$[-\hat{e}_{x}+\hat{e}_{y}]/\sqrt{2}$, while
it preserves the magnetization and $J_y$.
Thus,  $\sigma_{212}$ and $\sigma_{221}$ are forbidden by symmetry.

\section{Results}
\label{sec_results}
In this section we discuss the UMR and the NLHE 
in the ferromagnetic Rashba model introduced in
Sec.~\ref{sec_rashba_model}.
We set the mass $m^{*}$ in the Rashba model to the electron mass $m_{\rm e}$, i.e.,
$m^{*}=m_{\rm e}$.
A Rashba parameter 
of $\alpha^{\rm R}=0.095$eV\AA~\cite{sot_dmi_stiles}
has been estimated in
Co/Pt~\cite{sot_dmi_stiles} magnetic bilayers.
Very high $\alpha^{\rm R}$ parameters (up to $\alpha^{\rm R}=3.05$eV\AA) have been reported for
Bi/Ag(111) surface
alloys~\cite{PhysRevB.93.125409,giant_spin_splitting_surface_alloying}.
An even higher value of $\alpha^{\rm
  R}=3.85$eV\AA\,
has been reported for BiTeI~\cite{giant_rashba_spin_splitting_BiTeI}.
Our choice of $\alpha^{\rm R}$ in the numerical calculations below
covers a similar range of Rashba parameters.

\begin{figure}
\includegraphics[width=\linewidth]{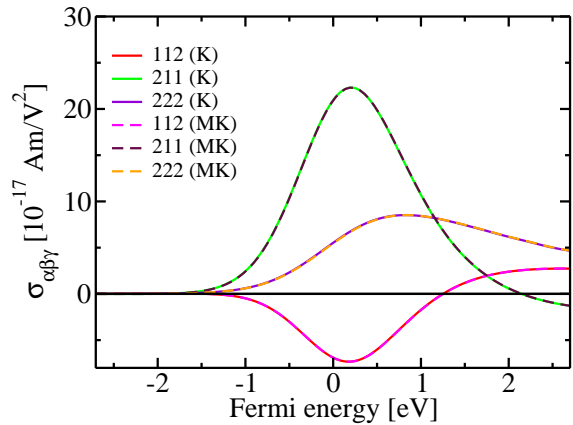}
\caption{\label{fig_compare_KM}
Nonlinear conductivity $\sigma_{\alpha\beta\gamma}$ vs.\ Fermi energy.
Comparison between the Keldysh approach (K) and the
Moyal-Keldysh approach (MK). $\alpha^{\rm R}=2$~eV\AA, $\Delta V=1$~eV, 
$\Gamma=1.36$~eV, and $\hat{\vn{M}}\Vert -\hat{\vn{e}}_{x}$
are used in the calculation. 
Both approaches yield identical results.
}
\end{figure}

Fig.~\ref{fig_compare_KM} shows the comparison between the
Keldysh and the Moyal-Keldysh approaches 
for the parameters $\alpha^{\rm R}=2$~eV\AA, $\Delta V=1$~eV, and $\Gamma=1.36$~eV
when the magnetization
points in the $-x$ direction. 
The figure demonstates that the
Keldysh and the Moyal-Keldysh approaches yield identical results
for the nonlinear conductivity $\sigma_{\alpha\beta\gamma}$,
which
corroborates the validity of both approaches.
In agreement with the symmetry 
analysis in Sec.~\ref{sec_symmetry} the following tensor components 
are zero (not shown in
the figure):
$\sigma_{111}$, $\sigma_{122}$, $\sigma_{212}$, $\sigma_{221}$.
Moreover, symmetry dictates that $\sigma_{112}=\sigma_{121}$
(therefore,
we show only $\sigma_{112}$ in the figure). 
When we compare the maxima of the UMR and the NLHE we find that
they are comparable in magnitude. When we investigate the
dependence of the UMR and of the NLHE on $\Gamma$ and on
$\alpha^{\rm R}$ in the figures below we find that this property persists
also when these parameters are changed.

\begin{figure}
\includegraphics[width=\linewidth]{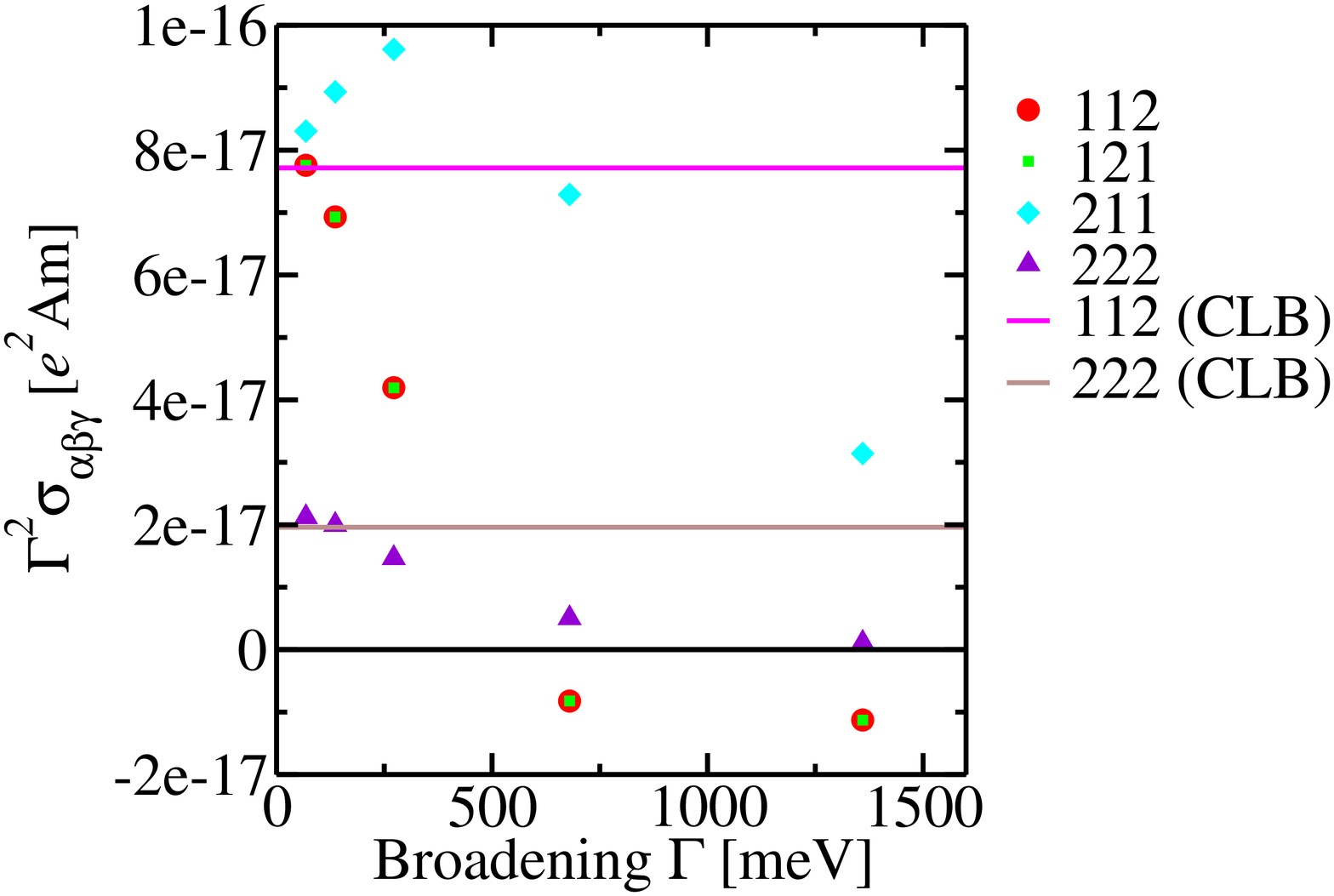}
\caption{\label{fig_scaling_with_gamma_fixfermi}
Tensor $\Gamma^2 \sigma_{\alpha\beta\gamma}$ vs.\
broadening $\Gamma$ 
when $\hat{\vn{M}}\Vert -\hat{\vn{e}}_{x}$, $\alpha^{\rm R}=720$~meV\AA,
$\mathcal{E}_{\rm F}=0$, and $\Delta V=1$~eV. Solid lines show the
results of the clean-limit Boltzmann (CLB) expression
Eq.~\eqref{eq_boltzmann}, while the results obtained 
from Eq.~\eqref{eq_keldysh_total}
are shown by symbols. CLB is symmetric under any permution of the
indices of $\sigma_{\alpha\beta\gamma}$. Consequently, we show only
112 (CLB) in the figure, because 211 (CLB) and 121 (CLB) are equal to it.
}
\end{figure}

In order to study the dependence of the UMR and of the NLHE
on the broadening $\Gamma$, we show the $\Gamma$-dependence
of the nonlinear conductivity in
Fig.~\ref{fig_scaling_with_gamma_fixfermi} at the Fermi energy $\mathcal{E}_{\rm
F}=0$, Rashba parameter $\alpha^{\rm R}=720$~meV\AA, and exchange splitting $\Delta V=1$~eV,
when $\hat{\vn{M}}$ points in the $-x$ direction.
In order to facilitate the illustration of the entire range from small
values of $\Gamma$ up to large values of $\Gamma$ we plot
$\Gamma^2 \sigma_{ijk}$ in this figure, because the factor $\Gamma^2$
compensates the $\propto \Gamma^{-2}$-behaviour expected in the clean
limit
according to Eq.~\eqref{eq_keldysh_intraband_ibp_gzero} 
and Eq.~\eqref{eq_keldysh_interband_ibp_gzero}.
The clean-limit Boltzmann result 
Eq.~\eqref{eq_final_boltzmann} is shown in the figure as well
by solid horizontal lines (CLB).
The figure shows that the deviations of  the clean-limit behaviour
from the complete Keldysh results
become
substantial when $\Gamma$ gets large. Such deviations might 
contribute to the discrepancies found between the Boltzmann-formalism
calculations and
the experiment in NiMnSb~\cite{2021unidirectionalNiMnSb}. 
Since the Boltzmann formalism yields a tensor $\sigma^{\rm(Boltz)}_{\alpha\beta\gamma}$
that is 
symmetric~\cite{zhang2021higherorder,deyo2009semiclassical,tsirkin2021separation,quantum_nonlin_hall_berry_curvature_dipole,2021unidirectionalNiMnSb} under permutation of the indices $\alpha$, $\beta$,
and $\gamma$, we show in Fig.~\ref{fig_scaling_with_gamma_fixfermi}
only the component $\sigma^{\rm(Boltz)}_{112}$
of the NLHE. In contrast, the Keldysh formalism predicts
$\sigma^{\rm(Boltz)}_{112}\ne \sigma^{\rm(Boltz)}_{211}$
when the clean-limit expression does not hold, i.e., when $\Gamma$ is
sufficiently large. Clearly, one may argue generally that the 
violation $\sigma^{\rm(Boltz)}_{112}\ne \sigma^{\rm(Boltz)}_{211}$
indicates that the relaxation time approximation within the
Boltzmann formalism fails on the quantitative level. 
Therefore,
when the violation $\sigma^{\rm(Boltz)}_{112}\ne
\sigma^{\rm(Boltz)}_{211}$ is established experimentally in a given
material,
one might consider this as an indication 
that one needs to go beyond the Boltzmann formalism with constant
relaxation time approximation to describe this effect theoretically.

\begin{figure}
\includegraphics[width=\linewidth]{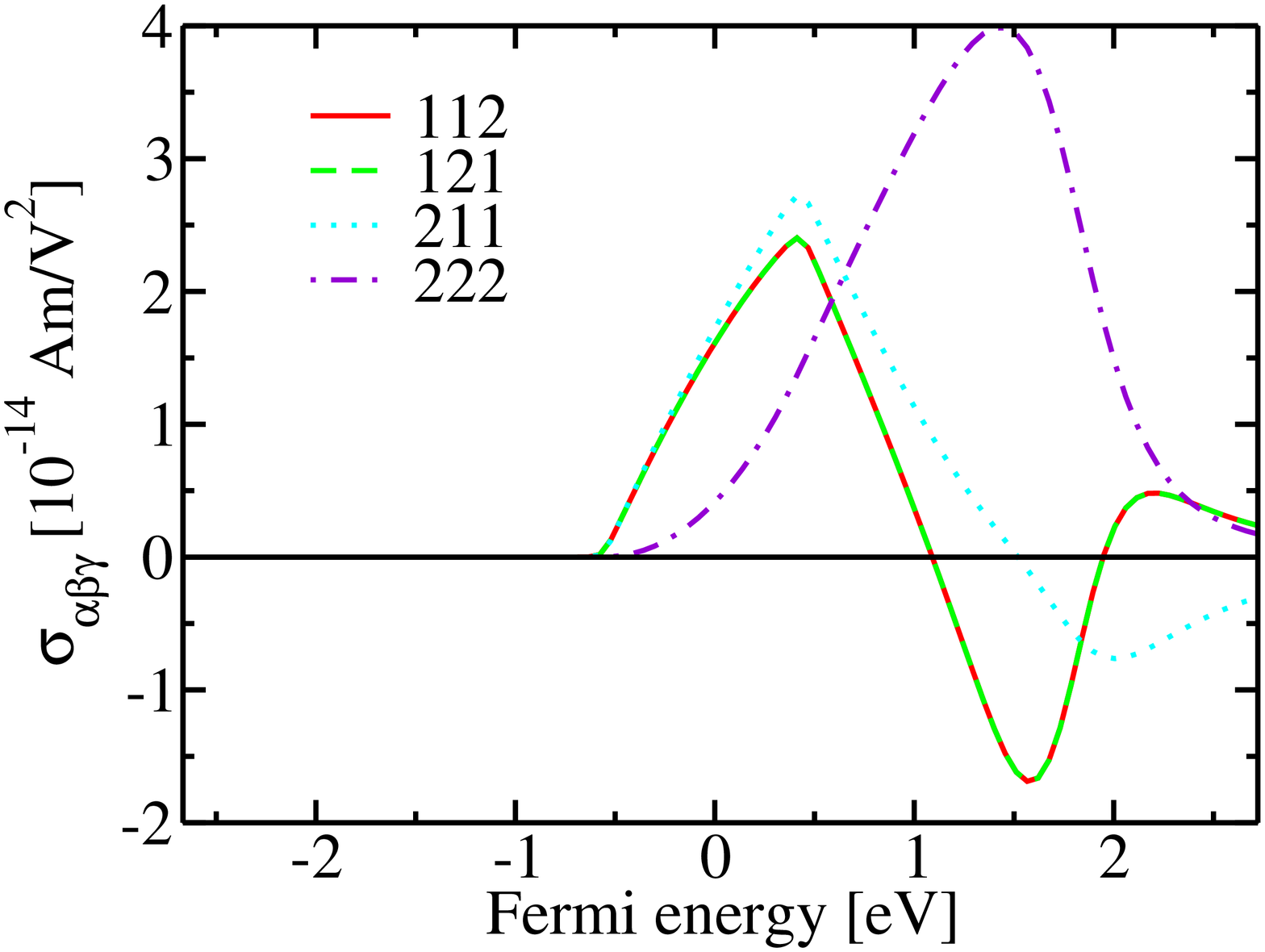}
\caption{\label{fig_sigma0005_alpha_0.1}
Nonlinear conductivity tensor $\sigma_{\alpha\beta\gamma}$ vs.\ Fermi energy 
when $\hat{\vn{M}}\Vert -\hat{\vn{e}}_{x}$, $\alpha^{\rm R}=720$~meV\AA, $\Delta V=1$~eV,  and $\Gamma=68$~meV. 
}
\end{figure}

\begin{figure}
\includegraphics[width=\linewidth]{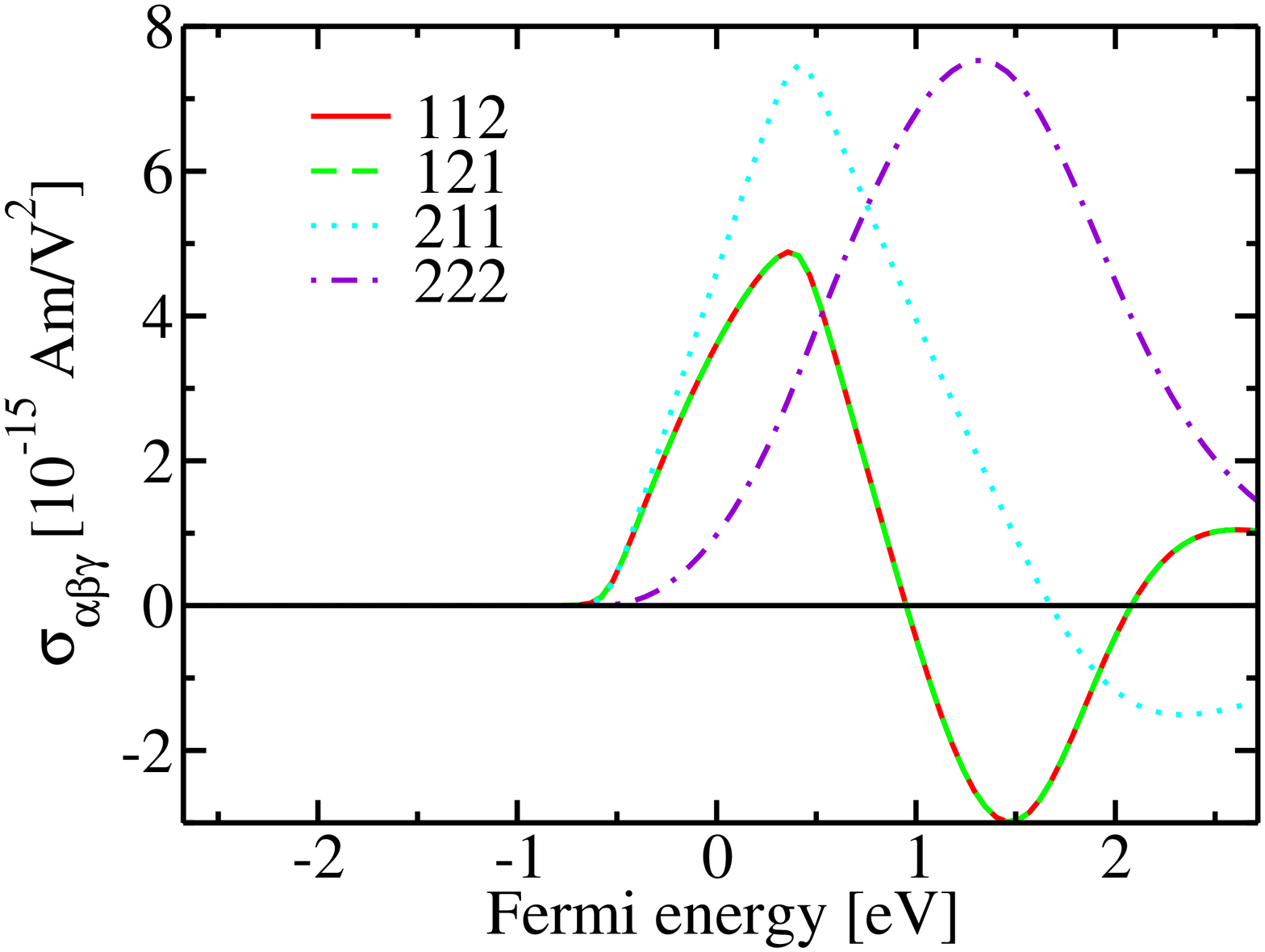}
\caption{\label{fig_sigma001_alpha_0.1}
Nonlinear conductivity tensor $\sigma_{\alpha\beta\gamma}$ vs.\ Fermi energy 
when $\hat{\vn{M}}\Vert -\hat{\vn{e}}_{x}$, $\alpha^{\rm R}=720$~meV\AA, $\Delta V=1$~eV,  and $\Gamma=136$~meV. 
}
\end{figure}

\begin{figure}
\includegraphics[width=\linewidth]{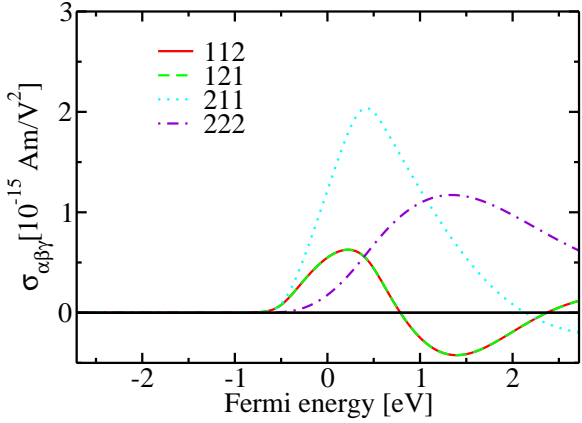}
\caption{\label{fig_sigma002_alpha_0.1}
Nonlinear conductivity tensor $\sigma_{\alpha\beta\gamma}$ vs.\ Fermi energy 
when $\hat{\vn{M}}\Vert -\hat{\vn{e}}_{x}$, $\alpha^{\rm R}=720$~meV\AA, $\Delta V=1$~eV,  and $\Gamma=272$~meV. 
}
\end{figure}

\begin{figure}
\includegraphics[width=\linewidth]{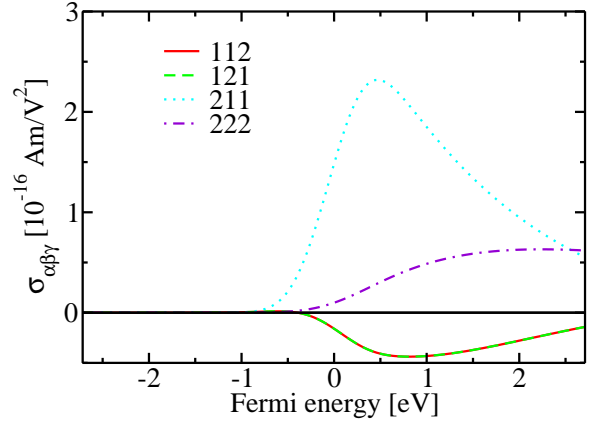}
\caption{\label{fig_sigma005_alpha_0.1}
Nonlinear conductivity tensor $\sigma_{\alpha\beta\gamma}$ vs.\ Fermi energy 
when $\hat{\vn{M}}\Vert -\hat{\vn{e}}_{x}$, $\alpha^{\rm R}=720$~meV\AA, $\Delta V=1$~eV,  and $\Gamma=680$~meV. 
}
\end{figure}

\begin{figure}
\includegraphics[width=\linewidth]{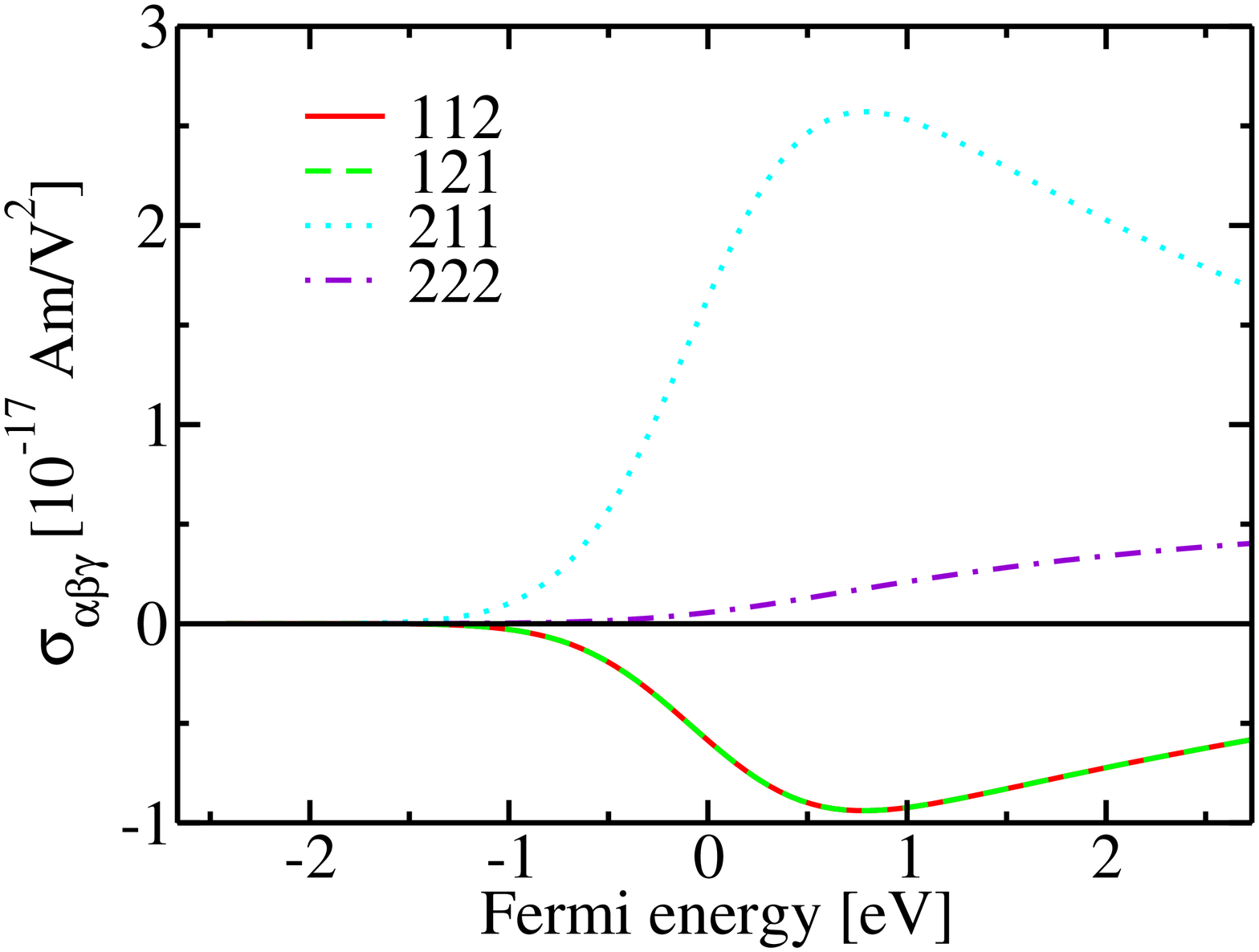}
\caption{\label{fig_big_sigma_alpha_0.1}
Nonlinear conductivity tensor $\sigma_{\alpha\beta\gamma}$ vs.\ Fermi energy 
when $\hat{\vn{M}}\Vert -\hat{\vn{e}}_{x}$, $\alpha^{\rm R}=720$~meV\AA, $\Delta V=1$~eV,  and $\Gamma=1.36$~eV. 
}
\end{figure}

In order to study the dependence of the UMR and of the NLHE
on the Fermi energy $\mathcal{E}_{\rm F}$ we show the 
nonlinear conductivity tensor  $\sigma_{\alpha\beta\gamma}$ as a function of
Fermi energy when the Rashba parameter is $\alpha^{\rm R}=720$~meV\AA\,
and when $\hat{\vn{M}}$ points in the $-x$ direction
for the broadenings
$\Gamma=68$~meV,
$\Gamma=136$~meV,
$\Gamma=272$~meV,
$\Gamma=680$~meV,
and
$\Gamma=1.36$~eV,
in  
Fig.~\ref{fig_sigma0005_alpha_0.1},
Fig.~\ref{fig_sigma001_alpha_0.1},
Fig.~\ref{fig_sigma002_alpha_0.1},
Fig.~\ref{fig_sigma005_alpha_0.1},
and
Fig.~\ref{fig_big_sigma_alpha_0.1},
respectively.
For these parameters the bandstructure of the Rashba model exhibits 
a crossing between the first band and the second band at around
1.84~eV, and the band minimum of the first band is at -0.53~eV, while
the
band minimum of the second band is at 0.47~eV. These band structure
properties are visible in Fig.~\ref{fig_sigma0005_alpha_0.1}:
The conductivities vanish below the band minimum of the first band,
where the density of states is zero. Around 1.84~eV, where the two
bands
cross, the conductivities exhibit maxima. The NLHE components exhibit
additional maxima around  0.47~eV, where the minimum of the second
band
is located.
These features start to change qualitatively if the broadening
$\Gamma$
 increases towards the scale of the energy spacing between these features. 
Since we discussed the $\Gamma$-dependence in
Fig.~\ref{fig_scaling_with_gamma_fixfermi}
only based on a single Fermi energy $\mathcal{E}_{\rm F}$,
we discuss it now a second time by comparing 
Fig.~\ref{fig_sigma0005_alpha_0.1} through
Fig.~\ref{fig_big_sigma_alpha_0.1}
in order to see if qualitative features such as maxima, minima and
zeros
in the curves are modified by $\Gamma$. 
As discussed in Fig.~\ref{fig_scaling_with_gamma_fixfermi}
we expect that $\sigma_{\alpha\beta\gamma}\propto \tau^2 \propto \Gamma^{-2}$
when $\Gamma$ is small. Based on this scaling we expect an
increase of $\sigma_{\alpha\beta\gamma}$ by a factor of 4 when going from
Fig.~\ref{fig_sigma001_alpha_0.1} to
Fig.~\ref{fig_sigma0005_alpha_0.1}. This expectation is roughly
satisfied and we attribute the deviations to the size of $\Gamma$,
which
is not small enough to yield the exact $\propto\Gamma^{-2}$ behaviour
of the clean limit.
At larger values of $\Gamma$ this $\propto\Gamma^{-2}$ rule becomes
less and less predictive. For example the curves in
Fig.~\ref{fig_sigma002_alpha_0.1} and  Fig.~\ref{fig_sigma005_alpha_0.1}
differ substantially qualitatively.

\begin{figure}
\includegraphics[width=\linewidth]{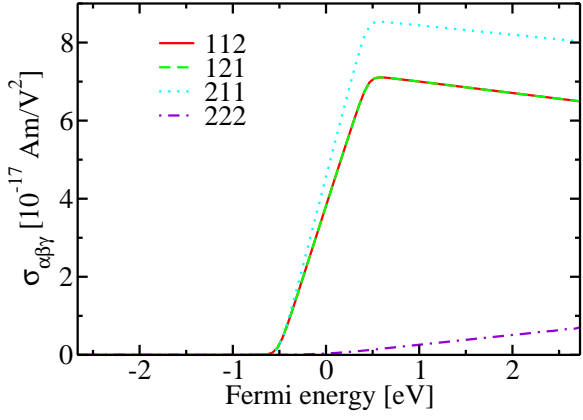}
\caption{\label{fig_integral_sigma001_alpha0.02}
Nonlinear conductivity tensor $\sigma_{\alpha\beta\gamma}$ vs.\ Fermi energy 
when $\hat{\vn{M}}\Vert -\hat{\vn{e}}_{x}$, $\alpha^{\rm R}=144$~meV\AA, $\Delta V=1$~eV,  and $\Gamma=136$~meV. 
}
\end{figure}

\begin{figure}
\includegraphics[width=\linewidth]{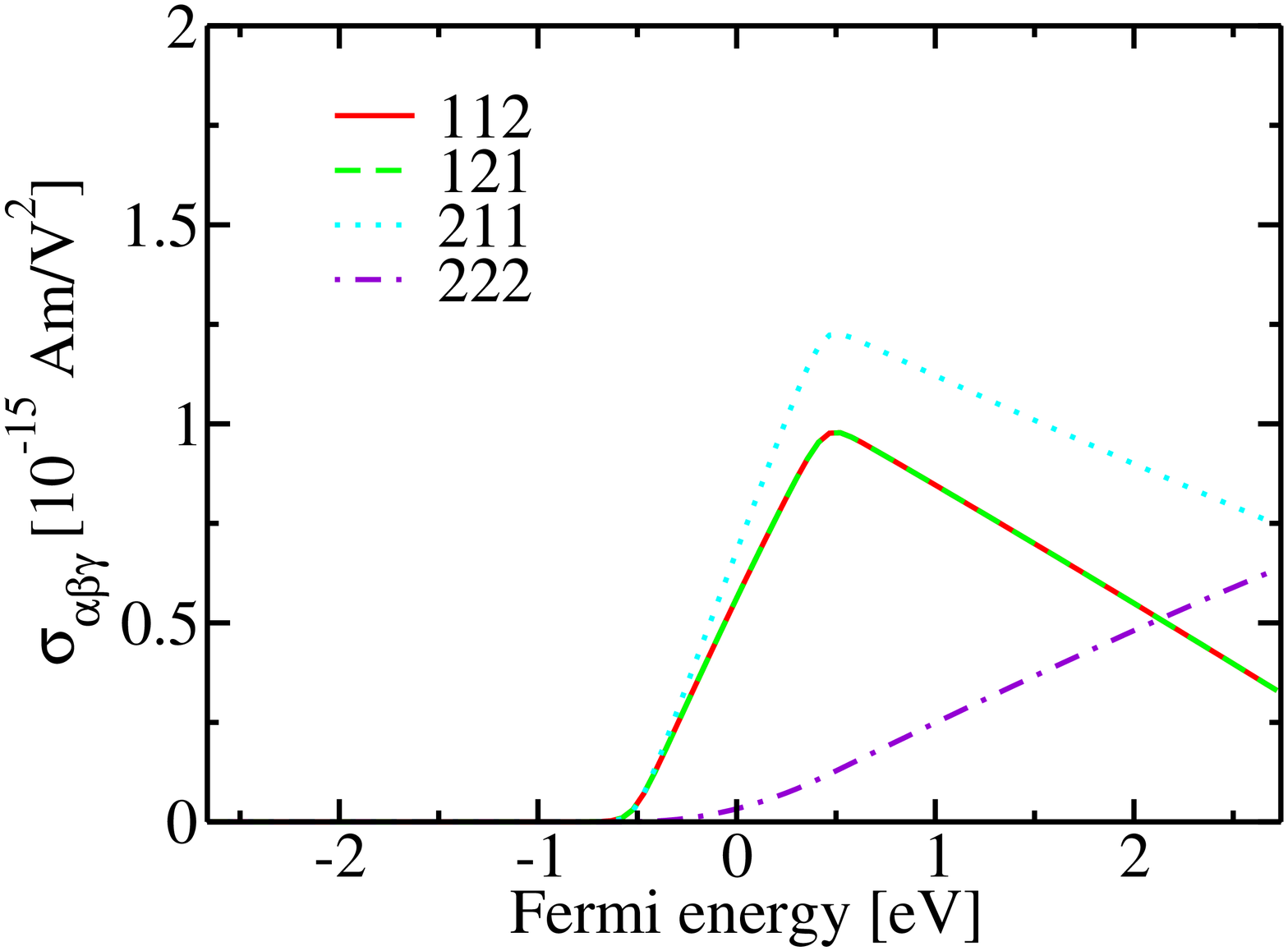}
\caption{\label{fig_integral_sigma001_alpha0.05}
Nonlinear conductivity tensor $\sigma_{\alpha\beta\gamma}$ vs.\ Fermi energy 
when $\hat{\vn{M}}\Vert -\hat{\vn{e}}_{x}$, $\alpha^{\rm R}=360$~meV\AA, $\Delta V=1$~eV,  and $\Gamma=136$~meV. 
}
\end{figure}

\begin{figure}
\includegraphics[width=\linewidth]{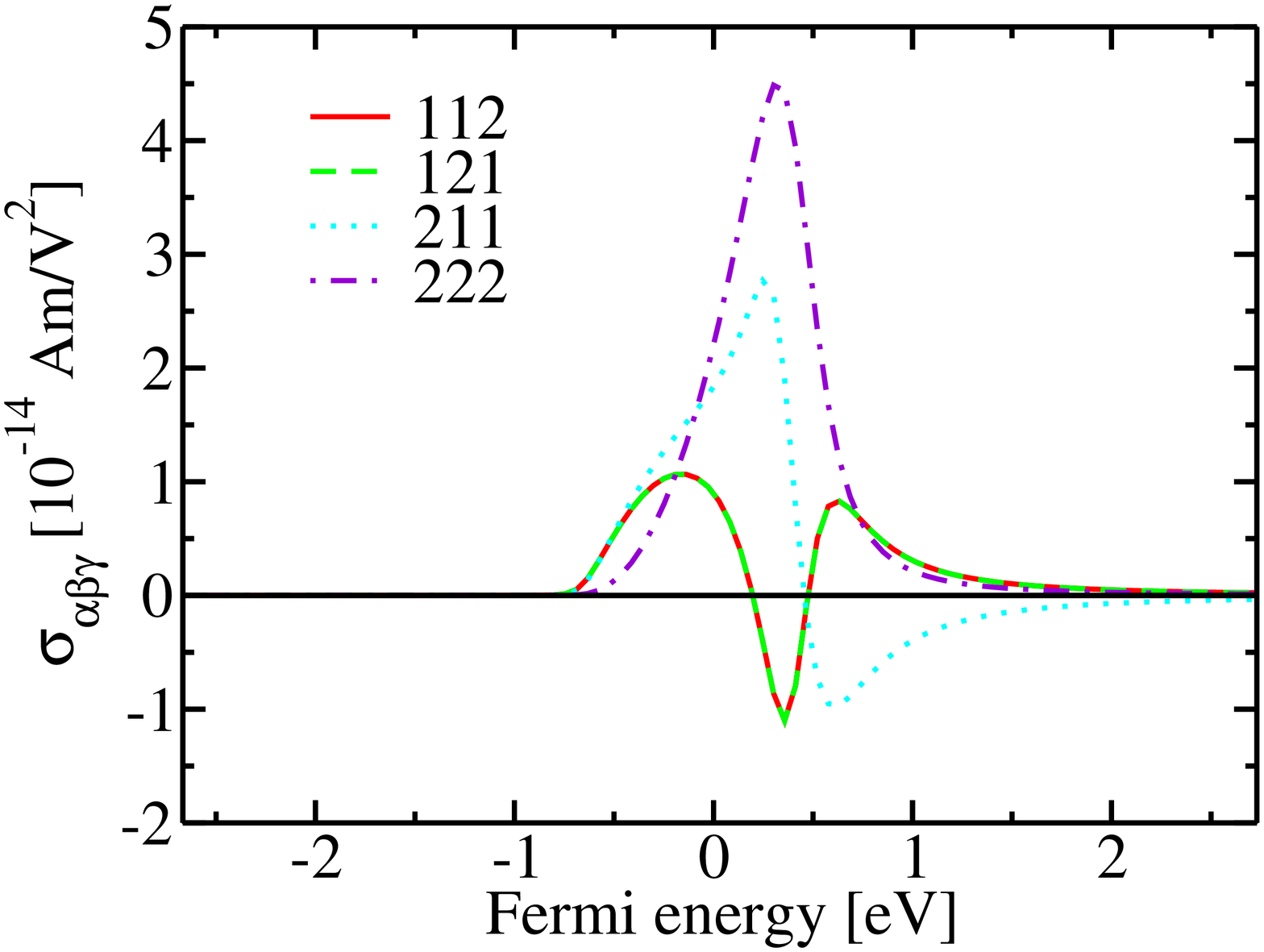}
\caption{\label{fig_integral_sigma001_alpha0.2}
Nonlinear conductivity tensor $\sigma_{\alpha\beta\gamma}$ vs.\ Fermi energy 
when $\hat{\vn{M}}\Vert -\hat{\vn{e}}_{x}$, $\alpha^{\rm R}=1439$~meV\AA, $\Delta V=1$~eV,  and $\Gamma=136$~meV. 
}
\end{figure}

\begin{figure}
\includegraphics[width=\linewidth]{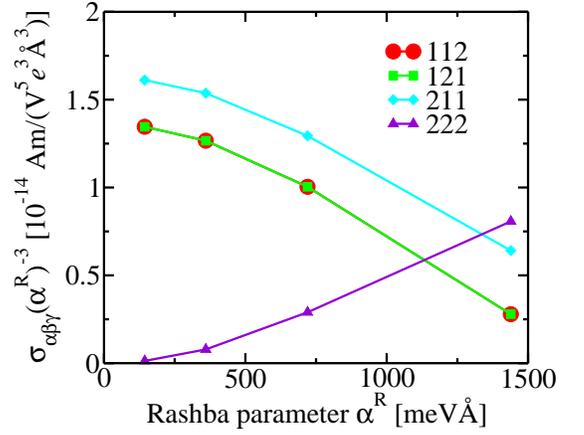}
\caption{\label{fig_sigma01_vs_alpha_collected_dividebyalpha3}
Tensor $\sigma_{\alpha\beta\gamma}/(\alpha^{\rm R})^3$ vs.\ Rashba
parameter $\alpha^{\rm R}$ 
when $\hat{\vn{M}}\Vert -\hat{\vn{e}}_{x}$, $\Delta V=1$~eV,
$\mathcal{E}_{\rm F}=0$ and
$\Gamma=136$~meV.
Results are shown by symbols and the solid lines only serve as guide
to the eye. 
}
\end{figure}

Next, we investigate the dependence on the Rashba parameter
$\alpha^{\rm R}$.
First, we fix the broadening to $\Gamma=136$~meV and vary
$\alpha^{\rm R}$.
Fig.~\ref{fig_integral_sigma001_alpha0.02},
Fig.~\ref{fig_integral_sigma001_alpha0.05},
Fig.~\ref{fig_sigma001_alpha_0.1}, and
Fig.~\ref{fig_integral_sigma001_alpha0.2}
show the nonlinear conductivity tensor for
$\alpha^{\rm R}=144$~meV\AA,
$\alpha^{\rm R}=360$~meV\AA,
$\alpha^{\rm R}=720$~meV\AA, and
$\alpha^{\rm R}=1439$~meV\AA,
respectively. Here, we observe that the maxima of $\sigma_{\alpha\beta\gamma}$
increase stronger than linearly with $\alpha^{\rm R}$.
Comparing for example Fig.~\ref{fig_sigma001_alpha_0.1} and
Fig.~\ref{fig_integral_sigma001_alpha0.2} we find that
$\sigma_{211}$ and $\sigma_{222}$ increase by roughly one order
of magnitude when $\alpha^{\rm R}$ is doubled.
In order to investigate this strong $\alpha^{\rm R}$-dependence in
more detail we plot the tensor
$\sigma_{\alpha\beta\gamma}/(\alpha^{\rm R})^3$
in Fig.~\ref{fig_sigma01_vs_alpha_collected_dividebyalpha3} for the
fixed Fermi energy of $\mathcal{E}_{\rm F}=0$.
The division of the nonlinear conductivity by the third power of the
Rashba parameter facilitates the illustration of the entire 
range of $\alpha^{\rm R}$ considered here. The component
$\sigma_{222}/(\alpha^{\rm R})^3$
depends roughly linearly on $\alpha^{\rm R}$ in the range considered
in the figure. Consequently,
in a coarse approximation $\sigma_{\alpha\beta\gamma}\propto
(\alpha^{\rm R})^4$ roughly predicts the trend in the range considered
in the figure.
In contrast, the NLHE depends less strongly on $\alpha^{\rm R}$ at
this
particular Fermi energy and consequently the components of
$\sigma_{\alpha\beta\gamma}/(\alpha^{\rm R})^3$
that correspond to the NLHE decrease with increasing $\alpha^{\rm R}$
in the range considered in the figure.

\begin{figure}
\includegraphics[width=\linewidth]{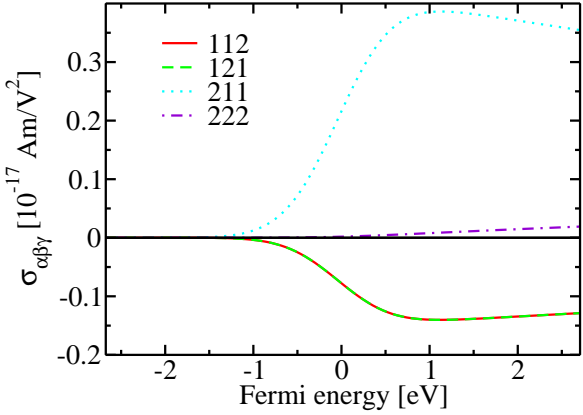}
\caption{\label{fig_big_sigma_alpha_0.05}
Nonlinear conductivity tensor $\sigma_{\alpha\beta\gamma}$ vs.\ Fermi energy 
when $\hat{\vn{M}}\Vert -\hat{\vn{e}}_{x}$, $\alpha^{\rm R}=360$~meV\AA, $\Delta V=1$~eV,  and $\Gamma=1.36$~eV. 
}
\end{figure}

\begin{figure}
\includegraphics[width=\linewidth]{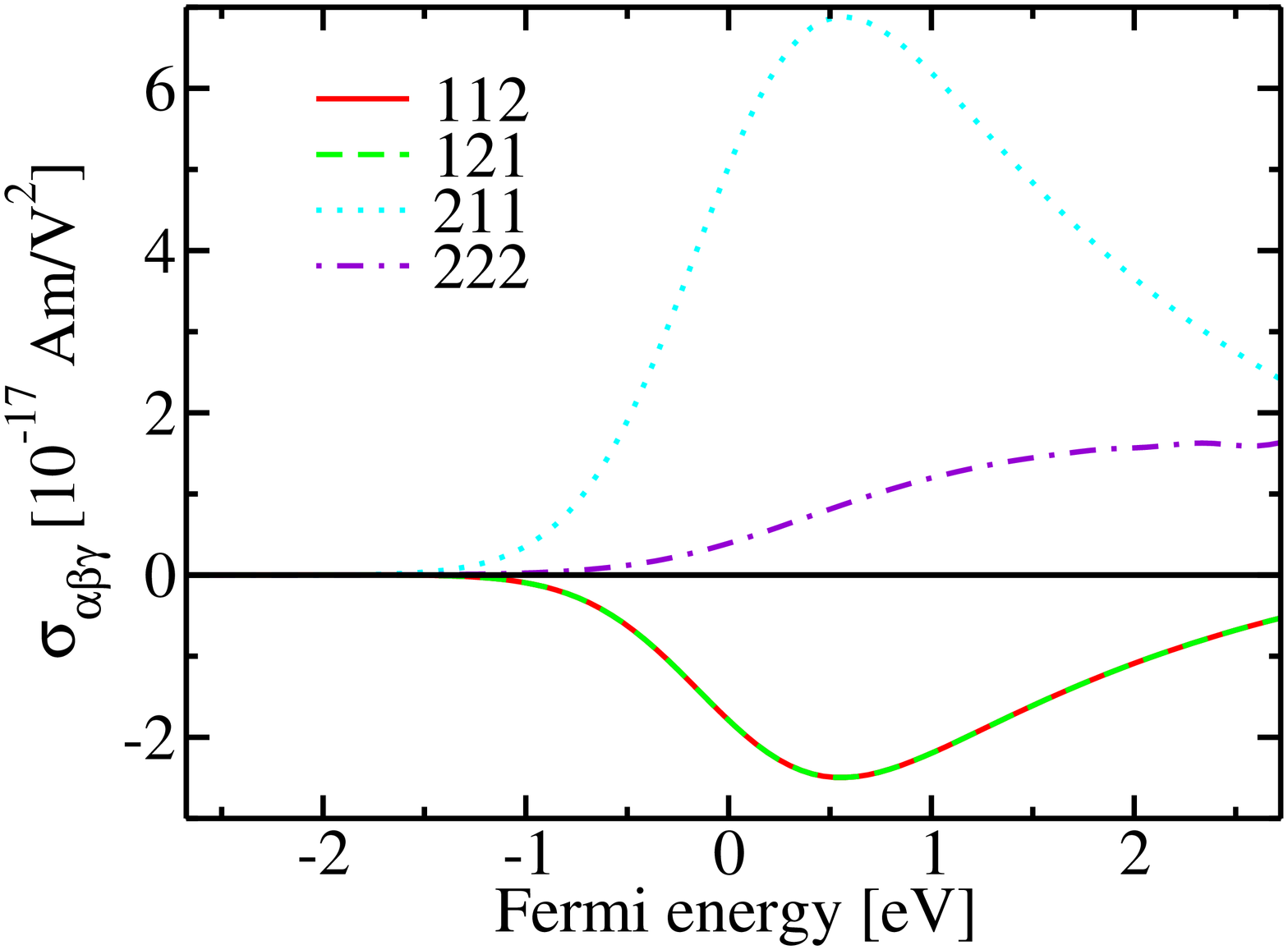}
\caption{\label{fig_big_sigma_alpha_0.15}
Nonlinear conductivity tensor $\sigma_{\alpha\beta\gamma}$ vs.\ Fermi energy 
when $\hat{\vn{M}}\Vert -\hat{\vn{e}}_{x}$, $\alpha^{\rm R}=1080$~meV\AA, $\Delta V=1$~eV,  and $\Gamma=1.36$~eV. 
}
\end{figure}

\begin{figure}
\includegraphics[width=\linewidth]{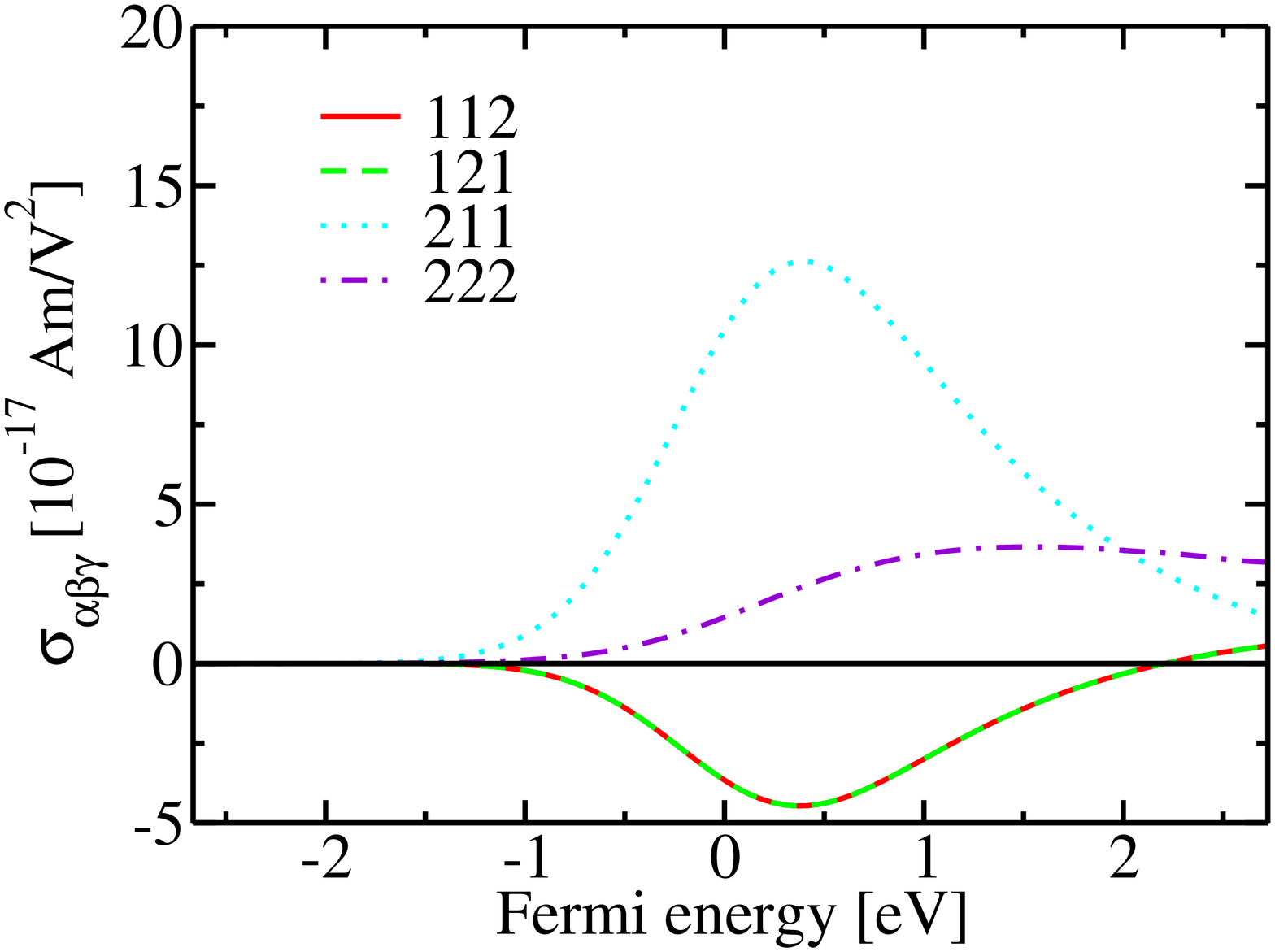}
\caption{\label{fig_big_sigma_alpha_0.2}  
Nonlinear conductivity tensor $\sigma_{\alpha\beta\gamma}$ vs.\ Fermi energy 
when $\hat{\vn{M}}\Vert -\hat{\vn{e}}_{x}$, $\alpha^{\rm R}=1439$~meV\AA, $\Delta V=1$~eV,  and $\Gamma=1.36$~eV. 
}
\end{figure}

\begin{figure}
\includegraphics[width=\linewidth]{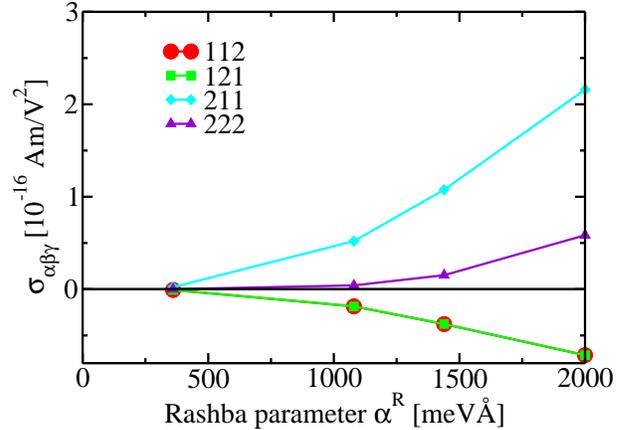}
\caption{\label{fig_sigma0.1_vs_alpha_collect_integrated}
Nonlinear conductivity $\sigma_{\alpha\beta\gamma}$ vs.\ Rashba
parameter $\alpha^{\rm R}$ 
when $\hat{\vn{M}}\Vert -\hat{\vn{e}}_{x}$, $\Delta V=1$~eV,
$\mathcal{E}_{\rm F}=0$ and
$\Gamma=1.36$~eV.
Results are shown by symbols and the solid lines only serve as guide
to the eye. 
}
\end{figure}

Finally, we study the dependence on the Rashba parameter $\alpha^{\rm R}$ at 
large broadening $\Gamma$ and we set $\Gamma=1.36$~eV.
We show the nonlinear conductivities for
$\alpha^{\rm R}=360$~meV\AA,
$\alpha^{\rm R}=1080$~meV\AA,
$\alpha^{\rm R}=1439$~meV\AA, and
$\alpha^{\rm R}=2$~eV\AA\,
in 
Fig.~\ref{fig_big_sigma_alpha_0.05},
Fig.~\ref{fig_big_sigma_alpha_0.15},
Fig.~\ref{fig_big_sigma_alpha_0.2}, and
Fig.~\ref{fig_compare_KM}.
When we compare Fig.~\ref{fig_big_sigma_alpha_0.2} and
Fig.~\ref{fig_compare_KM} we observe that $\sigma_{\alpha\beta\gamma}$
still increases stronger than linearly with $\alpha^{\rm R}$. However,
it does not increase as strongly as for $\Gamma=136$~meV
discussed in the preceding paragraph.
This behaviour is also illustrated in
Fig.~\ref{fig_sigma0.1_vs_alpha_collect_integrated},
which shows the nonlinear conductivity as a function of the Rashba
parameter
$\alpha^{\rm R}$ when the Fermi energy is set to $\mathcal{E}_{\rm F}=0$.

It is instructive to compare the magnitude of the UMR and NLHE
currents to the magnitude of the photocurrents induced at optical
frequencies. According to Ref.~\cite{lasincucspira} laser pulses 
with 1.55~eV photon energy and intensity 10GW/cm$^2$ induce
photocurrent densities of the order of magnitude of A/m for Rashba parameters
 similar to those considered here. The intensity 10GW/cm$^2$
corresponds to the electric field strength of the laser field 
of 2.7~MV/cm. When this field strength induces a photocurrent of 
1~A/m,
the corresponding second order response coefficient is
$\sigma_{\alpha\beta\gamma}=$~A/m (2.7~MV/cm)$^{-2}$=$1.37\times 10^{-17}$~Am/V$^2$.
This is the same order of magnitude as the response coefficients shown
in  Fig.~\ref{fig_big_sigma_alpha_0.05}, i.e., at very large
broadening of $1.36$~eV. In contrast, we find a response that is
larger by three orders of magnitude at small broadening
$\Gamma=68$~meV
shown in Fig.~\ref{fig_sigma0005_alpha_0.1}.
Since the calculations in Ref.~\cite{lasincucspira}
used small broadenings, we may conclude that the nonlinear
conductivity
at zero frequency is several orders of magnitude larger than the one
at optical frequencies when the broadenings are comparable.
This finding is consistent with the strong increase
and in some cases even divergent behaviour of the
quadratic response coefficients as $\omega\rightarrow 0$
found in studies of the inverse Faraday effect~\cite{PhysRevLett.117.137203}
and of the photocurrent~\cite{PhysRevX.10.041041,PhysRevB.97.241118,PhysRevB.102.121111}.

\section{Summary}
\label{sec_summary}
We derive the quadratic response of the electric current to an applied
electric field using two different formalisms: The usual Keldysh
nonequilibrium
formalism and the Moyal-Keldysh formalism. The latter approach
solves the difficulty of the non-periodic scalar potential associated
with a spatially homogeneous time-independent electric field
elegantly through the Moyal product. In contrast, the former approach
considers a spatially homogeneous time-dependent electric field
instead,
which may be described by the vector potential, and the zero-frequency
limit needs to be taken at the end of the derivation. We show that
these
two rather different approaches lead to numerically identical results
in the
ferromagnetic Rashba model, which corroborates their applicability
to magnetic Hamiltonians with SOI.
Since the Moyal-Keldysh formalism yields the zero-frequency dc
response directly, it is presumably the most convenient approach
 for non-linear response coefficients of high order, because taking
 the
zero-frequency limit in the Keldysh approach becomes 
more complex as the order of the perturbation increases.
When taking the zero-frequency limit in the Keldysh approach we
observe
that the second order dc conductivity is not identical to the
zero-frequency limit of the dc photocurrent expression, because the
zero-frequency limit of the
2nd harmonic generation contributes to the second order dc
conductivity as well. 
Additionally, we compare our Keldysh expression in the clean limit analytically to the literature
result
obtained from the Boltzmann formalism in the constant relaxation time
approximation, and find both formulae to agree in this limit.
We apply our quadratic response expressions to the ferromagnetic
Rashba model in order to study UMR and NLHE.
We find the UMR and the NLHE to be of comparable magnitude in this
model.
Additionally, they scale similarly with the Rashba strength and the
quasiparticle broadening. Compared to the response tensor that
describes the photocurrent generation at optical frequencies, the
zero-frequency effects considered here are several orders of magnitude
larger
when the parameters in the Rashba parameter are chosen similarly.
Our quadratic response expressions are also well-suited to study
UMR and NLHE within a first-principles density-functional theory framework.

\section*{Acknowledgments}We acknowledge financial support from
Leibniz Collaborative Excellence project OptiSPIN $-$ Optical Control
of Nanoscale Spin Textures, funding  under SPP 2137 ``Skyrmionics" of
the DFG
and Sino-German research project DISTOMAT (DFG project MO \mbox{1731/10-1}).
We gratefully acknowledge financial support from the European Research
Council (ERC) under
the European Union's Horizon 2020 research and innovation
program (Grant No. 856538, project ``3D MAGiC'').
The work was also supported by the Deutsche Forschungsgemeinschaft
(DFG, German Research Foundation) $-$ TRR 173 $-$ 268565370 (project
A11), TRR 288 $-$ 422213477 (project B06).  We  also gratefully
acknowledge the J\"ulich
Supercomputing Centre and RWTH Aachen University for providing
computational
resources under project No. jiff40.

\bibliography{secorddcre}

\begin{thebibliography}{41}%
\makeatletter
\providecommand \@ifxundefined [1]{%
 \@ifx{#1\undefined}
}%
\providecommand \@ifnum [1]{%
 \ifnum #1\expandafter \@firstoftwo
 \else \expandafter \@secondoftwo
 \fi
}%
\providecommand \@ifx [1]{%
 \ifx #1\expandafter \@firstoftwo
 \else \expandafter \@secondoftwo
 \fi
}%
\providecommand \natexlab [1]{#1}%
\providecommand \enquote  [1]{``#1''}%
\providecommand \bibnamefont  [1]{#1}%
\providecommand \bibfnamefont [1]{#1}%
\providecommand \citenamefont [1]{#1}%
\providecommand \href@noop [0]{\@secondoftwo}%
\providecommand \href [0]{\begingroup \@sanitize@url \@href}%
\providecommand \@href[1]{\@@startlink{#1}\@@href}%
\providecommand \@@href[1]{\endgroup#1\@@endlink}%
\providecommand \@sanitize@url [0]{\catcode `\\12\catcode `\$12\catcode
  `\&12\catcode `\#12\catcode `\^12\catcode `\_12\catcode `\%12\relax}%
\providecommand \@@startlink[1]{}%
\providecommand \@@endlink[0]{}%
\providecommand \url  [0]{\begingroup\@sanitize@url \@url }%
\providecommand \@url [1]{\endgroup\@href {#1}{\urlprefix }}%
\providecommand \urlprefix  [0]{URL }%
\providecommand \Eprint [0]{\href }%
\providecommand \doibase [0]{https://doi.org/}%
\providecommand \selectlanguage [0]{\@gobble}%
\providecommand \bibinfo  [0]{\@secondoftwo}%
\providecommand \bibfield  [0]{\@secondoftwo}%
\providecommand \translation [1]{[#1]}%
\providecommand \BibitemOpen [0]{}%
\providecommand \bibitemStop [0]{}%
\providecommand \bibitemNoStop [0]{.\EOS\space}%
\providecommand \EOS [0]{\spacefactor3000\relax}%
\providecommand \BibitemShut  [1]{\csname bibitem#1\endcsname}%
\let\auto@bib@innerbib\@empty
\bibitem [{\citenamefont {Avci}\ \emph
  {et~al.}(2015{\natexlab{a}})\citenamefont {Avci}, \citenamefont {Garello},
  \citenamefont {Mendil}, \citenamefont {Ghosh}, \citenamefont {Blasakis},
  \citenamefont {Gabureac}, \citenamefont {Trassin}, \citenamefont {Fiebig},\
  and\ \citenamefont {Gambardella}}]{usmr_bilayers_avci}%
  \BibitemOpen
  \bibfield  {author} {\bibinfo {author} {\bibfnamefont {C.~O.}\ \bibnamefont
  {Avci}}, \bibinfo {author} {\bibfnamefont {K.}~\bibnamefont {Garello}},
  \bibinfo {author} {\bibfnamefont {J.}~\bibnamefont {Mendil}}, \bibinfo
  {author} {\bibfnamefont {A.}~\bibnamefont {Ghosh}}, \bibinfo {author}
  {\bibfnamefont {N.}~\bibnamefont {Blasakis}}, \bibinfo {author}
  {\bibfnamefont {M.}~\bibnamefont {Gabureac}}, \bibinfo {author}
  {\bibfnamefont {M.}~\bibnamefont {Trassin}}, \bibinfo {author} {\bibfnamefont
  {M.}~\bibnamefont {Fiebig}},\ and\ \bibinfo {author} {\bibfnamefont
  {P.}~\bibnamefont {Gambardella}},\ }\bibfield  {title} {\bibinfo {title}
  {{Magnetoresistance of heavy and light metal/ferromagnet bilayers}},\
  }\href@noop {} {\bibfield  {journal} {\bibinfo  {journal} {{APPLIED PHYSICS
  LETTERS}}\ }\textbf {\bibinfo {volume} {{107}}},\ \bibinfo {pages} {{192405}}
  (\bibinfo {year} {{2015}}{\natexlab{a}})}\BibitemShut {NoStop}%
\bibitem [{\citenamefont {Avci}\ \emph
  {et~al.}(2015{\natexlab{b}})\citenamefont {Avci}, \citenamefont {Garello},
  \citenamefont {Ghosh}, \citenamefont {Gabureac}, \citenamefont {Alvarado},\
  and\ \citenamefont {Gambardella}}]{unidirectiona_Avci}%
  \BibitemOpen
  \bibfield  {author} {\bibinfo {author} {\bibfnamefont {C.~O.}\ \bibnamefont
  {Avci}}, \bibinfo {author} {\bibfnamefont {K.}~\bibnamefont {Garello}},
  \bibinfo {author} {\bibfnamefont {A.}~\bibnamefont {Ghosh}}, \bibinfo
  {author} {\bibfnamefont {M.}~\bibnamefont {Gabureac}}, \bibinfo {author}
  {\bibfnamefont {S.~F.}\ \bibnamefont {Alvarado}},\ and\ \bibinfo {author}
  {\bibfnamefont {P.}~\bibnamefont {Gambardella}},\ }\bibfield  {title}
  {\bibinfo {title} {{U}nidirectional spin {H}all magnetoresistance in
  ferromagnet/normal metal bilayers},\ }\href
  {https://doi.org/10.1038/nphys3356} {\bibfield  {journal} {\bibinfo
  {journal} {Nature physics}\ }\textbf {\bibinfo {volume} {11}},\ \bibinfo
  {pages} {570} (\bibinfo {year} {2015}{\natexlab{b}})}\BibitemShut {NoStop}%
\bibitem [{\citenamefont {Kim}\ \emph {et~al.}(2016)\citenamefont {Kim},
  \citenamefont {Sheng}, \citenamefont {Takahashi}, \citenamefont {Mitani},\
  and\ \citenamefont {Hayashi}}]{shmr_metallic_bilayers_hayashi}%
  \BibitemOpen
  \bibfield  {author} {\bibinfo {author} {\bibfnamefont {J.}~\bibnamefont
  {Kim}}, \bibinfo {author} {\bibfnamefont {P.}~\bibnamefont {Sheng}}, \bibinfo
  {author} {\bibfnamefont {S.}~\bibnamefont {Takahashi}}, \bibinfo {author}
  {\bibfnamefont {S.}~\bibnamefont {Mitani}},\ and\ \bibinfo {author}
  {\bibfnamefont {M.}~\bibnamefont {Hayashi}},\ }\bibfield  {title} {\bibinfo
  {title} {Spin hall magnetoresistance in metallic bilayers},\ }\href
  {https://doi.org/10.1103/PhysRevLett.116.097201} {\bibfield  {journal}
  {\bibinfo  {journal} {Phys. Rev. Lett.}\ }\textbf {\bibinfo {volume} {116}},\
  \bibinfo {pages} {097201} (\bibinfo {year} {2016})}\BibitemShut {NoStop}%
\bibitem [{\citenamefont {Olejn\'{\i}k}\ \emph {et~al.}(2015)\citenamefont
  {Olejn\'{\i}k}, \citenamefont {Nov\'ak}, \citenamefont {Wunderlich},\ and\
  \citenamefont {Jungwirth}}]{electrical_detection_magnetization_reversal}%
  \BibitemOpen
  \bibfield  {author} {\bibinfo {author} {\bibfnamefont {K.}~\bibnamefont
  {Olejn\'{\i}k}}, \bibinfo {author} {\bibfnamefont {V.}~\bibnamefont
  {Nov\'ak}}, \bibinfo {author} {\bibfnamefont {J.}~\bibnamefont
  {Wunderlich}},\ and\ \bibinfo {author} {\bibfnamefont {T.}~\bibnamefont
  {Jungwirth}},\ }\bibfield  {title} {\bibinfo {title} {Electrical detection of
  magnetization reversal without auxiliary magnets},\ }\href
  {https://doi.org/10.1103/PhysRevB.91.180402} {\bibfield  {journal} {\bibinfo
  {journal} {Phys. Rev. B}\ }\textbf {\bibinfo {volume} {91}},\ \bibinfo
  {pages} {180402(R)} (\bibinfo {year} {2015})}\BibitemShut {NoStop}%
\bibitem [{\citenamefont {Avci}\ \emph {et~al.}(2017)\citenamefont {Avci},
  \citenamefont {Mann}, \citenamefont {Tan}, \citenamefont {Gambardella},\ and\
  \citenamefont {Beach}}]{multi_state_memory_device_usmr}%
  \BibitemOpen
  \bibfield  {author} {\bibinfo {author} {\bibfnamefont {C.~O.}\ \bibnamefont
  {Avci}}, \bibinfo {author} {\bibfnamefont {M.}~\bibnamefont {Mann}}, \bibinfo
  {author} {\bibfnamefont {A.~J.}\ \bibnamefont {Tan}}, \bibinfo {author}
  {\bibfnamefont {P.}~\bibnamefont {Gambardella}},\ and\ \bibinfo {author}
  {\bibfnamefont {G.~S.~D.}\ \bibnamefont {Beach}},\ }\bibfield  {title}
  {\bibinfo {title} {A multi-state memory device based on the unidirectional
  spin hall magnetoresistance},\ }\href {https://doi.org/10.1063/1.4983784}
  {\bibfield  {journal} {\bibinfo  {journal} {Applied Physics Letters}\
  }\textbf {\bibinfo {volume} {110}},\ \bibinfo {pages} {203506} (\bibinfo
  {year} {2017})}\BibitemShut {NoStop}%
\bibitem [{\citenamefont {Zhang}\ and\ \citenamefont
  {Vignale}(2016)}]{unidirectional_theory}%
  \BibitemOpen
  \bibfield  {author} {\bibinfo {author} {\bibfnamefont {S.~S.-L.}\
  \bibnamefont {Zhang}}\ and\ \bibinfo {author} {\bibfnamefont
  {G.}~\bibnamefont {Vignale}},\ }\bibfield  {title} {\bibinfo {title} {Theory
  of unidirectional spin hall magnetoresistance in
  heavy-metal/ferromagnetic-metal bilayers},\ }\href
  {https://doi.org/10.1103/PhysRevB.94.140411} {\bibfield  {journal} {\bibinfo
  {journal} {Phys. Rev. B}\ }\textbf {\bibinfo {volume} {94}},\ \bibinfo
  {pages} {140411(R)} (\bibinfo {year} {2016})}\BibitemShut {NoStop}%
\bibitem [{\citenamefont {Avci}\ \emph {et~al.}(2018)\citenamefont {Avci},
  \citenamefont {Mendil}, \citenamefont {Beach},\ and\ \citenamefont
  {Gambardella}}]{origins_unidirectional_spin_hall_magnetoresistance}%
  \BibitemOpen
  \bibfield  {author} {\bibinfo {author} {\bibfnamefont {C.~O.}\ \bibnamefont
  {Avci}}, \bibinfo {author} {\bibfnamefont {J.}~\bibnamefont {Mendil}},
  \bibinfo {author} {\bibfnamefont {G.~S.~D.}\ \bibnamefont {Beach}},\ and\
  \bibinfo {author} {\bibfnamefont {P.}~\bibnamefont {Gambardella}},\
  }\bibfield  {title} {\bibinfo {title} {Origins of the unidirectional spin
  hall magnetoresistance in metallic bilayers},\ }\href
  {https://doi.org/10.1103/PhysRevLett.121.087207} {\bibfield  {journal}
  {\bibinfo  {journal} {Phys. Rev. Lett.}\ }\textbf {\bibinfo {volume} {121}},\
  \bibinfo {pages} {087207} (\bibinfo {year} {2018})}\BibitemShut {NoStop}%
\bibitem [{\citenamefont {Yin}\ \emph {et~al.}(2017)\citenamefont {Yin},
  \citenamefont {Han}, \citenamefont {de~Jong}, \citenamefont {Lavrijsen},
  \citenamefont {Duine}, \citenamefont {Swagten},\ and\ \citenamefont
  {Koopmans}}]{thickness_dependence_unidirectional_SHM}%
  \BibitemOpen
  \bibfield  {author} {\bibinfo {author} {\bibfnamefont {Y.}~\bibnamefont
  {Yin}}, \bibinfo {author} {\bibfnamefont {D.-S.}\ \bibnamefont {Han}},
  \bibinfo {author} {\bibfnamefont {M.~C.~H.}\ \bibnamefont {de~Jong}},
  \bibinfo {author} {\bibfnamefont {R.}~\bibnamefont {Lavrijsen}}, \bibinfo
  {author} {\bibfnamefont {R.~A.}\ \bibnamefont {Duine}}, \bibinfo {author}
  {\bibfnamefont {H.~J.~M.}\ \bibnamefont {Swagten}},\ and\ \bibinfo {author}
  {\bibfnamefont {B.}~\bibnamefont {Koopmans}},\ }\bibfield  {title} {\bibinfo
  {title} {Thickness dependence of unidirectional spin-hall magnetoresistance
  in metallic bilayers},\ }\href {https://doi.org/10.1063/1.5003725} {\bibfield
   {journal} {\bibinfo  {journal} {Applied Physics Letters}\ }\textbf {\bibinfo
  {volume} {111}},\ \bibinfo {pages} {232405} (\bibinfo {year}
  {2017})}\BibitemShut {NoStop}%
\bibitem [{\citenamefont {Guillet}\ \emph {et~al.}(2021)\citenamefont
  {Guillet}, \citenamefont {Marty}, \citenamefont {Vergnaud}, \citenamefont
  {Jamet}, \citenamefont {Zucchetti}, \citenamefont {Isella}, \citenamefont
  {Barbedienne}, \citenamefont {Jaffr\`es}, \citenamefont {Reyren},
  \citenamefont {George},\ and\ \citenamefont {Fert}}]{PhysRevB.103.064411}%
  \BibitemOpen
  \bibfield  {author} {\bibinfo {author} {\bibfnamefont {T.}~\bibnamefont
  {Guillet}}, \bibinfo {author} {\bibfnamefont {A.}~\bibnamefont {Marty}},
  \bibinfo {author} {\bibfnamefont {C.}~\bibnamefont {Vergnaud}}, \bibinfo
  {author} {\bibfnamefont {M.}~\bibnamefont {Jamet}}, \bibinfo {author}
  {\bibfnamefont {C.}~\bibnamefont {Zucchetti}}, \bibinfo {author}
  {\bibfnamefont {G.}~\bibnamefont {Isella}}, \bibinfo {author} {\bibfnamefont
  {Q.}~\bibnamefont {Barbedienne}}, \bibinfo {author} {\bibfnamefont
  {H.}~\bibnamefont {Jaffr\`es}}, \bibinfo {author} {\bibfnamefont
  {N.}~\bibnamefont {Reyren}}, \bibinfo {author} {\bibfnamefont {J.-M.}\
  \bibnamefont {George}},\ and\ \bibinfo {author} {\bibfnamefont
  {A.}~\bibnamefont {Fert}},\ }\bibfield  {title} {\bibinfo {title} {Large
  rashba unidirectional magnetoresistance in the fe/ge(111) interface states},\
  }\href {https://doi.org/10.1103/PhysRevB.103.064411} {\bibfield  {journal}
  {\bibinfo  {journal} {Phys. Rev. B}\ }\textbf {\bibinfo {volume} {103}},\
  \bibinfo {pages} {064411} (\bibinfo {year} {2021})}\BibitemShut {NoStop}%
\bibitem [{\citenamefont {Lv}\ \emph {et~al.}(2018)\citenamefont {Lv},
  \citenamefont {Kally}, \citenamefont {Zhang}, \citenamefont {Lee},
  \citenamefont {Jamali}, \citenamefont {Samarth},\ and\ \citenamefont
  {Wang}}]{unidirectional_spin_hall_and_rashba_edelstein}%
  \BibitemOpen
  \bibfield  {author} {\bibinfo {author} {\bibfnamefont {Y.}~\bibnamefont
  {Lv}}, \bibinfo {author} {\bibfnamefont {J.}~\bibnamefont {Kally}}, \bibinfo
  {author} {\bibfnamefont {D.}~\bibnamefont {Zhang}}, \bibinfo {author}
  {\bibfnamefont {J.~S.}\ \bibnamefont {Lee}}, \bibinfo {author} {\bibfnamefont
  {M.}~\bibnamefont {Jamali}}, \bibinfo {author} {\bibfnamefont
  {N.}~\bibnamefont {Samarth}},\ and\ \bibinfo {author} {\bibfnamefont {J.-P.}\
  \bibnamefont {Wang}},\ }\bibfield  {title} {\bibinfo {title} {{Unidirectional
  spin-Hall and Rashba-Edelstein magnetoresistance in topological
  insulator-ferromagnet layer heterostructures}},\ }\href@noop {} {\bibfield
  {journal} {\bibinfo  {journal} {{NATURE COMMUNICATIONS}}\ }\textbf {\bibinfo
  {volume} {{9}}},\ \bibinfo {pages} {{111}} (\bibinfo {year}
  {{2018}})}\BibitemShut {NoStop}%
\bibitem [{\citenamefont {Yasuda}\ \emph {et~al.}(2016)\citenamefont {Yasuda},
  \citenamefont {Tsukazaki}, \citenamefont {Yoshimi}, \citenamefont
  {Takahashi}, \citenamefont {Kawasaki},\ and\ \citenamefont
  {Tokura}}]{large_unidirectional_mr_magnetic_ti}%
  \BibitemOpen
  \bibfield  {author} {\bibinfo {author} {\bibfnamefont {K.}~\bibnamefont
  {Yasuda}}, \bibinfo {author} {\bibfnamefont {A.}~\bibnamefont {Tsukazaki}},
  \bibinfo {author} {\bibfnamefont {R.}~\bibnamefont {Yoshimi}}, \bibinfo
  {author} {\bibfnamefont {K.~S.}\ \bibnamefont {Takahashi}}, \bibinfo {author}
  {\bibfnamefont {M.}~\bibnamefont {Kawasaki}},\ and\ \bibinfo {author}
  {\bibfnamefont {Y.}~\bibnamefont {Tokura}},\ }\bibfield  {title} {\bibinfo
  {title} {Large unidirectional magnetoresistance in a magnetic topological
  insulator},\ }\href {https://doi.org/10.1103/PhysRevLett.117.127202}
  {\bibfield  {journal} {\bibinfo  {journal} {Phys. Rev. Lett.}\ }\textbf
  {\bibinfo {volume} {117}},\ \bibinfo {pages} {127202} (\bibinfo {year}
  {2016})}\BibitemShut {NoStop}%
\bibitem [{\citenamefont {Železný}\ \emph {et~al.}(2021)\citenamefont
  {Železný}, \citenamefont {Fang}, \citenamefont {Olejník}, \citenamefont
  {Patchett}, \citenamefont {Gerhard}, \citenamefont {Gould}, \citenamefont
  {Molenkamp}, \citenamefont {Gomez-Olivella}, \citenamefont {Zemen},
  \citenamefont {Tichý}, \citenamefont {Jungwirth},\ and\ \citenamefont
  {Ciccarelli}}]{2021unidirectionalNiMnSb}%
  \BibitemOpen
  \bibfield  {author} {\bibinfo {author} {\bibfnamefont {J.}~\bibnamefont
  {Železný}}, \bibinfo {author} {\bibfnamefont {Z.}~\bibnamefont {Fang}},
  \bibinfo {author} {\bibfnamefont {K.}~\bibnamefont {Olejník}}, \bibinfo
  {author} {\bibfnamefont {J.}~\bibnamefont {Patchett}}, \bibinfo {author}
  {\bibfnamefont {F.}~\bibnamefont {Gerhard}}, \bibinfo {author} {\bibfnamefont
  {C.}~\bibnamefont {Gould}}, \bibinfo {author} {\bibfnamefont {L.~W.}\
  \bibnamefont {Molenkamp}}, \bibinfo {author} {\bibfnamefont {C.}~\bibnamefont
  {Gomez-Olivella}}, \bibinfo {author} {\bibfnamefont {J.}~\bibnamefont
  {Zemen}}, \bibinfo {author} {\bibfnamefont {T.}~\bibnamefont {Tichý}},
  \bibinfo {author} {\bibfnamefont {T.}~\bibnamefont {Jungwirth}},\ and\
  \bibinfo {author} {\bibfnamefont {C.}~\bibnamefont {Ciccarelli}},\
  }\href@noop {} {\bibinfo {title} {Unidirectional magnetoresistance and
  spin-orbit torque in $\mathrm{NiMnSb}$}} (\bibinfo {year} {2021}),\ \Eprint
  {https://arxiv.org/abs/2102.12838} {arXiv:2102.12838 [cond-mat.mes-hall]}
  \BibitemShut {NoStop}%
\bibitem [{\citenamefont {Sodemann}\ and\ \citenamefont
  {Fu}(2015)}]{quantum_nonlin_hall_berry_curvature_dipole}%
  \BibitemOpen
  \bibfield  {author} {\bibinfo {author} {\bibfnamefont {I.}~\bibnamefont
  {Sodemann}}\ and\ \bibinfo {author} {\bibfnamefont {L.}~\bibnamefont {Fu}},\
  }\bibfield  {title} {\bibinfo {title} {Quantum nonlinear hall effect induced
  by berry curvature dipole in time-reversal invariant materials},\ }\href
  {https://doi.org/10.1103/PhysRevLett.115.216806} {\bibfield  {journal}
  {\bibinfo  {journal} {Phys. Rev. Lett.}\ }\textbf {\bibinfo {volume} {115}},\
  \bibinfo {pages} {216806} (\bibinfo {year} {2015})}\BibitemShut {NoStop}%
\bibitem [{\citenamefont {Yasuda}\ \emph {et~al.}(2017)\citenamefont {Yasuda},
  \citenamefont {Tsukazaki}, \citenamefont {Yoshimi}, \citenamefont {Kondou},
  \citenamefont {Takahashi}, \citenamefont {Otani}, \citenamefont {Kawasaki},\
  and\ \citenamefont {Tokura}}]{current_nonlinear_hall_effect}%
  \BibitemOpen
  \bibfield  {author} {\bibinfo {author} {\bibfnamefont {K.}~\bibnamefont
  {Yasuda}}, \bibinfo {author} {\bibfnamefont {A.}~\bibnamefont {Tsukazaki}},
  \bibinfo {author} {\bibfnamefont {R.}~\bibnamefont {Yoshimi}}, \bibinfo
  {author} {\bibfnamefont {K.}~\bibnamefont {Kondou}}, \bibinfo {author}
  {\bibfnamefont {K.~S.}\ \bibnamefont {Takahashi}}, \bibinfo {author}
  {\bibfnamefont {Y.}~\bibnamefont {Otani}}, \bibinfo {author} {\bibfnamefont
  {M.}~\bibnamefont {Kawasaki}},\ and\ \bibinfo {author} {\bibfnamefont
  {Y.}~\bibnamefont {Tokura}},\ }\bibfield  {title} {\bibinfo {title}
  {Current-nonlinear hall effect and spin-orbit torque magnetization switching
  in a magnetic topological insulator},\ }\href
  {https://doi.org/10.1103/PhysRevLett.119.137204} {\bibfield  {journal}
  {\bibinfo  {journal} {Phys. Rev. Lett.}\ }\textbf {\bibinfo {volume} {119}},\
  \bibinfo {pages} {137204} (\bibinfo {year} {2017})}\BibitemShut {NoStop}%
\bibitem [{\citenamefont {Sterk}\ \emph {et~al.}(2019)\citenamefont {Sterk},
  \citenamefont {Peerlings},\ and\ \citenamefont {Duine}}]{PhysRevB.99.064438}%
  \BibitemOpen
  \bibfield  {author} {\bibinfo {author} {\bibfnamefont {W.~P.}\ \bibnamefont
  {Sterk}}, \bibinfo {author} {\bibfnamefont {D.}~\bibnamefont {Peerlings}},\
  and\ \bibinfo {author} {\bibfnamefont {R.~A.}\ \bibnamefont {Duine}},\
  }\bibfield  {title} {\bibinfo {title} {Magnon contribution to unidirectional
  spin hall magnetoresistance in ferromagnetic-insulator/heavy-metal
  bilayers},\ }\href {https://doi.org/10.1103/PhysRevB.99.064438} {\bibfield
  {journal} {\bibinfo  {journal} {Phys. Rev. B}\ }\textbf {\bibinfo {volume}
  {99}},\ \bibinfo {pages} {064438} (\bibinfo {year} {2019})}\BibitemShut
  {NoStop}%
\bibitem [{\citenamefont {Watanabe}\ and\ \citenamefont
  {Yanase}(2020)}]{PhysRevResearch.2.043081}%
  \BibitemOpen
  \bibfield  {author} {\bibinfo {author} {\bibfnamefont {H.}~\bibnamefont
  {Watanabe}}\ and\ \bibinfo {author} {\bibfnamefont {Y.}~\bibnamefont
  {Yanase}},\ }\bibfield  {title} {\bibinfo {title} {Nonlinear electric
  transport in odd-parity magnetic multipole systems: Application to mn-based
  compounds},\ }\href {https://doi.org/10.1103/PhysRevResearch.2.043081}
  {\bibfield  {journal} {\bibinfo  {journal} {Phys. Rev. Research}\ }\textbf
  {\bibinfo {volume} {2}},\ \bibinfo {pages} {043081} (\bibinfo {year}
  {2020})}\BibitemShut {NoStop}%
\bibitem [{\citenamefont {Sipe}\ and\ \citenamefont
  {Shkrebtii}(2000)}]{PhysRevB.61.5337}%
  \BibitemOpen
  \bibfield  {author} {\bibinfo {author} {\bibfnamefont {J.~E.}\ \bibnamefont
  {Sipe}}\ and\ \bibinfo {author} {\bibfnamefont {A.~I.}\ \bibnamefont
  {Shkrebtii}},\ }\bibfield  {title} {\bibinfo {title} {Second-order optical
  response in semiconductors},\ }\href
  {https://doi.org/10.1103/PhysRevB.61.5337} {\bibfield  {journal} {\bibinfo
  {journal} {Phys. Rev. B}\ }\textbf {\bibinfo {volume} {61}},\ \bibinfo
  {pages} {5337} (\bibinfo {year} {2000})}\BibitemShut {NoStop}%
\bibitem [{\citenamefont {Taguchi}\ \emph
  {et~al.}(2016{\natexlab{a}})\citenamefont {Taguchi}, \citenamefont {Xu},
  \citenamefont {Yamakage},\ and\ \citenamefont {Law}}]{PhysRevB.94.155206}%
  \BibitemOpen
  \bibfield  {author} {\bibinfo {author} {\bibfnamefont {K.}~\bibnamefont
  {Taguchi}}, \bibinfo {author} {\bibfnamefont {D.-H.}\ \bibnamefont {Xu}},
  \bibinfo {author} {\bibfnamefont {A.}~\bibnamefont {Yamakage}},\ and\
  \bibinfo {author} {\bibfnamefont {K.~T.}\ \bibnamefont {Law}},\ }\bibfield
  {title} {\bibinfo {title} {Photovoltaic anomalous hall effect in line-node
  semimetals},\ }\href {https://doi.org/10.1103/PhysRevB.94.155206} {\bibfield
  {journal} {\bibinfo  {journal} {Phys. Rev. B}\ }\textbf {\bibinfo {volume}
  {94}},\ \bibinfo {pages} {155206} (\bibinfo {year}
  {2016}{\natexlab{a}})}\BibitemShut {NoStop}%
\bibitem [{\citenamefont {Taguchi}\ \emph
  {et~al.}(2016{\natexlab{b}})\citenamefont {Taguchi}, \citenamefont {Imaeda},
  \citenamefont {Sato},\ and\ \citenamefont {Tanaka}}]{PhysRevB.93.201202}%
  \BibitemOpen
  \bibfield  {author} {\bibinfo {author} {\bibfnamefont {K.}~\bibnamefont
  {Taguchi}}, \bibinfo {author} {\bibfnamefont {T.}~\bibnamefont {Imaeda}},
  \bibinfo {author} {\bibfnamefont {M.}~\bibnamefont {Sato}},\ and\ \bibinfo
  {author} {\bibfnamefont {Y.}~\bibnamefont {Tanaka}},\ }\bibfield  {title}
  {\bibinfo {title} {Photovoltaic chiral magnetic effect in weyl semimetals},\
  }\href {https://doi.org/10.1103/PhysRevB.93.201202} {\bibfield  {journal}
  {\bibinfo  {journal} {Phys. Rev. B}\ }\textbf {\bibinfo {volume} {93}},\
  \bibinfo {pages} {201202(R)} (\bibinfo {year}
  {2016}{\natexlab{b}})}\BibitemShut {NoStop}%
\bibitem [{\citenamefont {Ventura}\ \emph {et~al.}(2017)\citenamefont
  {Ventura}, \citenamefont {Passos}, \citenamefont {Lopes~dos Santos},
  \citenamefont {Viana Parente~Lopes},\ and\ \citenamefont
  {Peres}}]{PhysRevB.96.035431}%
  \BibitemOpen
  \bibfield  {author} {\bibinfo {author} {\bibfnamefont {G.~B.}\ \bibnamefont
  {Ventura}}, \bibinfo {author} {\bibfnamefont {D.~J.}\ \bibnamefont {Passos}},
  \bibinfo {author} {\bibfnamefont {J.~M.~B.}\ \bibnamefont {Lopes~dos
  Santos}}, \bibinfo {author} {\bibfnamefont {J.~M.}\ \bibnamefont {Viana
  Parente~Lopes}},\ and\ \bibinfo {author} {\bibfnamefont {N.~M.~R.}\
  \bibnamefont {Peres}},\ }\bibfield  {title} {\bibinfo {title} {Gauge
  covariances and nonlinear optical responses},\ }\href
  {https://doi.org/10.1103/PhysRevB.96.035431} {\bibfield  {journal} {\bibinfo
  {journal} {Phys. Rev. B}\ }\textbf {\bibinfo {volume} {96}},\ \bibinfo
  {pages} {035431} (\bibinfo {year} {2017})}\BibitemShut {NoStop}%
\bibitem [{\citenamefont {Taghizadeh}\ and\ \citenamefont
  {Pedersen}(2018)}]{PhysRevB.97.205432}%
  \BibitemOpen
  \bibfield  {author} {\bibinfo {author} {\bibfnamefont {A.}~\bibnamefont
  {Taghizadeh}}\ and\ \bibinfo {author} {\bibfnamefont {T.~G.}\ \bibnamefont
  {Pedersen}},\ }\bibfield  {title} {\bibinfo {title} {Gauge invariance of
  excitonic linear and nonlinear optical response},\ }\href
  {https://doi.org/10.1103/PhysRevB.97.205432} {\bibfield  {journal} {\bibinfo
  {journal} {Phys. Rev. B}\ }\textbf {\bibinfo {volume} {97}},\ \bibinfo
  {pages} {205432} (\bibinfo {year} {2018})}\BibitemShut {NoStop}%
\bibitem [{\citenamefont {Passos}\ \emph {et~al.}(2018)\citenamefont {Passos},
  \citenamefont {Ventura}, \citenamefont {Viana Parente~Lopes}, \citenamefont
  {Lopes~dos Santos},\ and\ \citenamefont {Peres}}]{PhysRevB.97.235446}%
  \BibitemOpen
  \bibfield  {author} {\bibinfo {author} {\bibfnamefont {D.~J.}\ \bibnamefont
  {Passos}}, \bibinfo {author} {\bibfnamefont {G.~B.}\ \bibnamefont {Ventura}},
  \bibinfo {author} {\bibfnamefont {J.~M.}\ \bibnamefont {Viana
  Parente~Lopes}}, \bibinfo {author} {\bibfnamefont {J.~M.~B.}\ \bibnamefont
  {Lopes~dos Santos}},\ and\ \bibinfo {author} {\bibfnamefont {N.~M.~R.}\
  \bibnamefont {Peres}},\ }\bibfield  {title} {\bibinfo {title} {Nonlinear
  optical responses of crystalline systems: Results from a velocity gauge
  analysis},\ }\href {https://doi.org/10.1103/PhysRevB.97.235446} {\bibfield
  {journal} {\bibinfo  {journal} {Phys. Rev. B}\ }\textbf {\bibinfo {volume}
  {97}},\ \bibinfo {pages} {235446} (\bibinfo {year} {2018})}\BibitemShut
  {NoStop}%
\bibitem [{\citenamefont {Parker}\ \emph {et~al.}(2019)\citenamefont {Parker},
  \citenamefont {Morimoto}, \citenamefont {Orenstein},\ and\ \citenamefont
  {Moore}}]{PhysRevB.99.045121}%
  \BibitemOpen
  \bibfield  {author} {\bibinfo {author} {\bibfnamefont {D.~E.}\ \bibnamefont
  {Parker}}, \bibinfo {author} {\bibfnamefont {T.}~\bibnamefont {Morimoto}},
  \bibinfo {author} {\bibfnamefont {J.}~\bibnamefont {Orenstein}},\ and\
  \bibinfo {author} {\bibfnamefont {J.~E.}\ \bibnamefont {Moore}},\ }\bibfield
  {title} {\bibinfo {title} {Diagrammatic approach to nonlinear optical
  response with application to weyl semimetals},\ }\href
  {https://doi.org/10.1103/PhysRevB.99.045121} {\bibfield  {journal} {\bibinfo
  {journal} {Phys. Rev. B}\ }\textbf {\bibinfo {volume} {99}},\ \bibinfo
  {pages} {045121} (\bibinfo {year} {2019})}\BibitemShut {NoStop}%
\bibitem [{\citenamefont {Jo{\~{a}}o}\ and\ \citenamefont
  {Lopes}(2019)}]{Jo_o_2019}%
  \BibitemOpen
  \bibfield  {author} {\bibinfo {author} {\bibfnamefont {S.~M.}\ \bibnamefont
  {Jo{\~{a}}o}}\ and\ \bibinfo {author} {\bibfnamefont {J.~M. V.~P.}\
  \bibnamefont {Lopes}},\ }\bibfield  {title} {\bibinfo {title}
  {Basis-independent spectral methods for non-linear optical response in
  arbitrary tight-binding models},\ }\href
  {https://doi.org/10.1088/1361-648x/ab59ec} {\bibfield  {journal} {\bibinfo
  {journal} {Journal of Physics: Condensed Matter}\ }\textbf {\bibinfo {volume}
  {32}},\ \bibinfo {pages} {125901} (\bibinfo {year} {2019})}\BibitemShut
  {NoStop}%
\bibitem [{\citenamefont {Freimuth}\ \emph {et~al.}(2016)\citenamefont
  {Freimuth}, \citenamefont {Bl\"ugel},\ and\ \citenamefont
  {Mokrousov}}]{lasintor}%
  \BibitemOpen
  \bibfield  {author} {\bibinfo {author} {\bibfnamefont {F.}~\bibnamefont
  {Freimuth}}, \bibinfo {author} {\bibfnamefont {S.}~\bibnamefont {Bl\"ugel}},\
  and\ \bibinfo {author} {\bibfnamefont {Y.}~\bibnamefont {Mokrousov}},\
  }\bibfield  {title} {\bibinfo {title} {Laser-induced torques in metallic
  ferromagnets},\ }\href@noop {} {\bibfield  {journal} {\bibinfo  {journal}
  {Phys. Rev. B}\ }\textbf {\bibinfo {volume} {94}},\ \bibinfo {pages} {144432}
  (\bibinfo {year} {2016})}\BibitemShut {NoStop}%
\bibitem [{\citenamefont {Freimuth}\ \emph {et~al.}(2021)\citenamefont
  {Freimuth}, \citenamefont {Bl\"ugel},\ and\ \citenamefont
  {Mokrousov}}]{lasincucspira}%
  \BibitemOpen
  \bibfield  {author} {\bibinfo {author} {\bibfnamefont {F.}~\bibnamefont
  {Freimuth}}, \bibinfo {author} {\bibfnamefont {S.}~\bibnamefont {Bl\"ugel}},\
  and\ \bibinfo {author} {\bibfnamefont {Y.}~\bibnamefont {Mokrousov}},\
  }\bibfield  {title} {\bibinfo {title} {Charge and spin photocurrents in the
  rashba model},\ }\href {https://doi.org/10.1103/PhysRevB.103.075428}
  {\bibfield  {journal} {\bibinfo  {journal} {Phys. Rev. B}\ }\textbf {\bibinfo
  {volume} {103}},\ \bibinfo {pages} {075428} (\bibinfo {year}
  {2021})}\BibitemShut {NoStop}%
\bibitem [{\citenamefont {Ahn}\ \emph {et~al.}(2020)\citenamefont {Ahn},
  \citenamefont {Guo},\ and\ \citenamefont {Nagaosa}}]{PhysRevX.10.041041}%
  \BibitemOpen
  \bibfield  {author} {\bibinfo {author} {\bibfnamefont {J.}~\bibnamefont
  {Ahn}}, \bibinfo {author} {\bibfnamefont {G.-Y.}\ \bibnamefont {Guo}},\ and\
  \bibinfo {author} {\bibfnamefont {N.}~\bibnamefont {Nagaosa}},\ }\bibfield
  {title} {\bibinfo {title} {Low-frequency divergence and quantum geometry of
  the bulk photovoltaic effect in topological semimetals},\ }\href
  {https://doi.org/10.1103/PhysRevX.10.041041} {\bibfield  {journal} {\bibinfo
  {journal} {Phys. Rev. X}\ }\textbf {\bibinfo {volume} {10}},\ \bibinfo
  {pages} {041041} (\bibinfo {year} {2020})}\BibitemShut {NoStop}%
\bibitem [{\citenamefont {Berritta}\ \emph {et~al.}(2016)\citenamefont
  {Berritta}, \citenamefont {Mondal}, \citenamefont {Carva},\ and\
  \citenamefont {Oppeneer}}]{PhysRevLett.117.137203}%
  \BibitemOpen
  \bibfield  {author} {\bibinfo {author} {\bibfnamefont {M.}~\bibnamefont
  {Berritta}}, \bibinfo {author} {\bibfnamefont {R.}~\bibnamefont {Mondal}},
  \bibinfo {author} {\bibfnamefont {K.}~\bibnamefont {Carva}},\ and\ \bibinfo
  {author} {\bibfnamefont {P.~M.}\ \bibnamefont {Oppeneer}},\ }\bibfield
  {title} {\bibinfo {title} {Ab initio theory of coherent laser-induced
  magnetization in metals},\ }\href
  {https://doi.org/10.1103/PhysRevLett.117.137203} {\bibfield  {journal}
  {\bibinfo  {journal} {Phys. Rev. Lett.}\ }\textbf {\bibinfo {volume} {117}},\
  \bibinfo {pages} {137203} (\bibinfo {year} {2016})}\BibitemShut {NoStop}%
\bibitem [{\citenamefont {Ogata}\ and\ \citenamefont
  {Fukuyama}(2015)}]{oms_ogata_fukuyama}%
  \BibitemOpen
  \bibfield  {author} {\bibinfo {author} {\bibfnamefont {M.}~\bibnamefont
  {Ogata}}\ and\ \bibinfo {author} {\bibfnamefont {H.}~\bibnamefont
  {Fukuyama}},\ }\bibfield  {title} {\bibinfo {title} {Orbital magnetism of
  bloch electrons i. general formula},\ }\href@noop {} {\bibfield  {journal}
  {\bibinfo  {journal} {Journal of the Physical Society of Japan}\ }\textbf
  {\bibinfo {volume} {84}},\ \bibinfo {pages} {124708} (\bibinfo {year}
  {2015})}\BibitemShut {NoStop}%
\bibitem [{\citenamefont {Onoda}\ \emph {et~al.}(2006)\citenamefont {Onoda},
  \citenamefont {Sugimoto},\ and\ \citenamefont
  {Nagaosa}}]{theory_noneq_const_emf}%
  \BibitemOpen
  \bibfield  {author} {\bibinfo {author} {\bibfnamefont {S.}~\bibnamefont
  {Onoda}}, \bibinfo {author} {\bibfnamefont {N.}~\bibnamefont {Sugimoto}},\
  and\ \bibinfo {author} {\bibfnamefont {N.}~\bibnamefont {Nagaosa}},\
  }\bibfield  {title} {\bibinfo {title} {{Theory of non-equilibirum states
  driven by constant electromagnetic fields - Non-commutative quantum mechanics
  in the Keldysh formalism}},\ }\href {https://doi.org/{10.1143/PTP.116.61}}
  {\bibfield  {journal} {\bibinfo  {journal} {{PROGRESS OF THEORETICAL
  PHYSICS}}\ }\textbf {\bibinfo {volume} {{116}}},\ \bibinfo {pages} {61}
  (\bibinfo {year} {{2006}})}\BibitemShut {NoStop}%
\bibitem [{\citenamefont {Freimuth}\ \emph {et~al.}(2013)\citenamefont
  {Freimuth}, \citenamefont {Bamler}, \citenamefont {Mokrousov},\ and\
  \citenamefont {Rosch}}]{phase_space_berry}%
  \BibitemOpen
  \bibfield  {author} {\bibinfo {author} {\bibfnamefont {F.}~\bibnamefont
  {Freimuth}}, \bibinfo {author} {\bibfnamefont {R.}~\bibnamefont {Bamler}},
  \bibinfo {author} {\bibfnamefont {Y.}~\bibnamefont {Mokrousov}},\ and\
  \bibinfo {author} {\bibfnamefont {A.}~\bibnamefont {Rosch}},\ }\bibfield
  {title} {\bibinfo {title} {Phase-space berry phases in chiral magnets:
  Dzyaloshinskii-moriya interaction and the charge of skyrmions},\ }\href@noop
  {} {\bibfield  {journal} {\bibinfo  {journal} {Phys. Rev. B}\ }\textbf
  {\bibinfo {volume} {88}},\ \bibinfo {pages} {214409} (\bibinfo {year}
  {2013})}\BibitemShut {NoStop}%
\bibitem [{\citenamefont {Manchon}\ \emph {et~al.}(2015)\citenamefont
  {Manchon}, \citenamefont {Koo}, \citenamefont {Nitta}, \citenamefont
  {Frolov},\ and\ \citenamefont {Duine}}]{rashba_review}%
  \BibitemOpen
  \bibfield  {author} {\bibinfo {author} {\bibfnamefont {A.}~\bibnamefont
  {Manchon}}, \bibinfo {author} {\bibfnamefont {H.~C.}\ \bibnamefont {Koo}},
  \bibinfo {author} {\bibfnamefont {J.}~\bibnamefont {Nitta}}, \bibinfo
  {author} {\bibfnamefont {S.~M.}\ \bibnamefont {Frolov}},\ and\ \bibinfo
  {author} {\bibfnamefont {R.~A.}\ \bibnamefont {Duine}},\ }\bibfield  {title}
  {\bibinfo {title} {{N}ew perspectives for {R}ashba spin–orbit coupling},\
  }\href@noop {} {\bibfield  {journal} {\bibinfo  {journal} {Nature materials}\
  }\textbf {\bibinfo {volume} {14}},\ \bibinfo {pages} {871} (\bibinfo {year}
  {2015})}\BibitemShut {NoStop}%
\bibitem [{\citenamefont {Carbone}\ \emph {et~al.}(2016)\citenamefont
  {Carbone}, \citenamefont {Moras}, \citenamefont {Sheverdyaeva}, \citenamefont
  {Pacil\'e}, \citenamefont {Papagno}, \citenamefont {Ferrari}, \citenamefont
  {Topwal}, \citenamefont {Vescovo}, \citenamefont {Bihlmayer}, \citenamefont
  {Freimuth}, \citenamefont {Mokrousov},\ and\ \citenamefont
  {Bl\"ugel}}]{PhysRevB.93.125409}%
  \BibitemOpen
  \bibfield  {author} {\bibinfo {author} {\bibfnamefont {C.}~\bibnamefont
  {Carbone}}, \bibinfo {author} {\bibfnamefont {P.}~\bibnamefont {Moras}},
  \bibinfo {author} {\bibfnamefont {P.~M.}\ \bibnamefont {Sheverdyaeva}},
  \bibinfo {author} {\bibfnamefont {D.}~\bibnamefont {Pacil\'e}}, \bibinfo
  {author} {\bibfnamefont {M.}~\bibnamefont {Papagno}}, \bibinfo {author}
  {\bibfnamefont {L.}~\bibnamefont {Ferrari}}, \bibinfo {author} {\bibfnamefont
  {D.}~\bibnamefont {Topwal}}, \bibinfo {author} {\bibfnamefont
  {E.}~\bibnamefont {Vescovo}}, \bibinfo {author} {\bibfnamefont
  {G.}~\bibnamefont {Bihlmayer}}, \bibinfo {author} {\bibfnamefont
  {F.}~\bibnamefont {Freimuth}}, \bibinfo {author} {\bibfnamefont
  {Y.}~\bibnamefont {Mokrousov}},\ and\ \bibinfo {author} {\bibfnamefont
  {S.}~\bibnamefont {Bl\"ugel}},\ }\bibfield  {title} {\bibinfo {title}
  {Asymmetric band gaps in a rashba film system},\ }\href
  {https://doi.org/10.1103/PhysRevB.93.125409} {\bibfield  {journal} {\bibinfo
  {journal} {Phys. Rev. B}\ }\textbf {\bibinfo {volume} {93}},\ \bibinfo
  {pages} {125409} (\bibinfo {year} {2016})}\BibitemShut {NoStop}%
\bibitem [{\citenamefont {Kim}\ \emph {et~al.}(2013)\citenamefont {Kim},
  \citenamefont {Lee}, \citenamefont {Lee},\ and\ \citenamefont
  {Stiles}}]{sot_dmi_stiles}%
  \BibitemOpen
  \bibfield  {author} {\bibinfo {author} {\bibfnamefont {K.-W.}\ \bibnamefont
  {Kim}}, \bibinfo {author} {\bibfnamefont {H.-W.}\ \bibnamefont {Lee}},
  \bibinfo {author} {\bibfnamefont {K.-J.}\ \bibnamefont {Lee}},\ and\ \bibinfo
  {author} {\bibfnamefont {M.~D.}\ \bibnamefont {Stiles}},\ }\bibfield  {title}
  {\bibinfo {title} {Chirality from interfacial spin-orbit coupling effects in
  magnetic bilayers},\ }\href {https://doi.org/10.1103/PhysRevLett.111.216601}
  {\bibfield  {journal} {\bibinfo  {journal} {Phys. Rev. Lett.}\ }\textbf
  {\bibinfo {volume} {111}},\ \bibinfo {pages} {216601} (\bibinfo {year}
  {2013})}\BibitemShut {NoStop}%
\bibitem [{\citenamefont {Ast}\ \emph {et~al.}(2007)\citenamefont {Ast},
  \citenamefont {Henk}, \citenamefont {Ernst}, \citenamefont {Moreschini},
  \citenamefont {Falub}, \citenamefont {Pacil\'e}, \citenamefont {Bruno},
  \citenamefont {Kern},\ and\ \citenamefont
  {Grioni}}]{giant_spin_splitting_surface_alloying}%
  \BibitemOpen
  \bibfield  {author} {\bibinfo {author} {\bibfnamefont {C.~R.}\ \bibnamefont
  {Ast}}, \bibinfo {author} {\bibfnamefont {J.}~\bibnamefont {Henk}}, \bibinfo
  {author} {\bibfnamefont {A.}~\bibnamefont {Ernst}}, \bibinfo {author}
  {\bibfnamefont {L.}~\bibnamefont {Moreschini}}, \bibinfo {author}
  {\bibfnamefont {M.~C.}\ \bibnamefont {Falub}}, \bibinfo {author}
  {\bibfnamefont {D.}~\bibnamefont {Pacil\'e}}, \bibinfo {author}
  {\bibfnamefont {P.}~\bibnamefont {Bruno}}, \bibinfo {author} {\bibfnamefont
  {K.}~\bibnamefont {Kern}},\ and\ \bibinfo {author} {\bibfnamefont
  {M.}~\bibnamefont {Grioni}},\ }\bibfield  {title} {\bibinfo {title} {Giant
  spin splitting through surface alloying},\ }\href
  {https://doi.org/10.1103/PhysRevLett.98.186807} {\bibfield  {journal}
  {\bibinfo  {journal} {Phys. Rev. Lett.}\ }\textbf {\bibinfo {volume} {98}},\
  \bibinfo {pages} {186807} (\bibinfo {year} {2007})}\BibitemShut {NoStop}%
\bibitem [{\citenamefont {Ishizaka}\ \emph {et~al.}(2011)\citenamefont
  {Ishizaka}, \citenamefont {Bahramy}, \citenamefont {Murakawa}, \citenamefont
  {Sakano}, \citenamefont {Shimojima}, \citenamefont {Sonobe}, \citenamefont
  {Koizumi}, \citenamefont {Shin}, \citenamefont {Miyahara}, \citenamefont
  {Kimura}, \citenamefont {Miyamoto}, \citenamefont {Okuda}, \citenamefont
  {Namatame}, \citenamefont {Taniguchi}, \citenamefont {Arita}, \citenamefont
  {Nagaosa}, \citenamefont {Kobayashi}, \citenamefont {Murakami}, \citenamefont
  {Kumai}, \citenamefont {Kaneko}, \citenamefont {Onose},\ and\ \citenamefont
  {Tokura}}]{giant_rashba_spin_splitting_BiTeI}%
  \BibitemOpen
  \bibfield  {author} {\bibinfo {author} {\bibfnamefont {K.}~\bibnamefont
  {Ishizaka}}, \bibinfo {author} {\bibfnamefont {M.~S.}\ \bibnamefont
  {Bahramy}}, \bibinfo {author} {\bibfnamefont {H.}~\bibnamefont {Murakawa}},
  \bibinfo {author} {\bibfnamefont {M.}~\bibnamefont {Sakano}}, \bibinfo
  {author} {\bibfnamefont {T.}~\bibnamefont {Shimojima}}, \bibinfo {author}
  {\bibfnamefont {T.}~\bibnamefont {Sonobe}}, \bibinfo {author} {\bibfnamefont
  {K.}~\bibnamefont {Koizumi}}, \bibinfo {author} {\bibfnamefont
  {S.}~\bibnamefont {Shin}}, \bibinfo {author} {\bibfnamefont {H.}~\bibnamefont
  {Miyahara}}, \bibinfo {author} {\bibfnamefont {A.}~\bibnamefont {Kimura}},
  \bibinfo {author} {\bibfnamefont {K.}~\bibnamefont {Miyamoto}}, \bibinfo
  {author} {\bibfnamefont {T.}~\bibnamefont {Okuda}}, \bibinfo {author}
  {\bibfnamefont {H.}~\bibnamefont {Namatame}}, \bibinfo {author}
  {\bibfnamefont {M.}~\bibnamefont {Taniguchi}}, \bibinfo {author}
  {\bibfnamefont {R.}~\bibnamefont {Arita}}, \bibinfo {author} {\bibfnamefont
  {N.}~\bibnamefont {Nagaosa}}, \bibinfo {author} {\bibfnamefont
  {K.}~\bibnamefont {Kobayashi}}, \bibinfo {author} {\bibfnamefont
  {Y.}~\bibnamefont {Murakami}}, \bibinfo {author} {\bibfnamefont
  {R.}~\bibnamefont {Kumai}}, \bibinfo {author} {\bibfnamefont
  {Y.}~\bibnamefont {Kaneko}}, \bibinfo {author} {\bibfnamefont
  {Y.}~\bibnamefont {Onose}},\ and\ \bibinfo {author} {\bibfnamefont
  {Y.}~\bibnamefont {Tokura}},\ }\bibfield  {title} {\bibinfo {title} {Giant
  rashba-type spin splitting in bulk $\mathrm{BiTeI}$},\ }\href
  {https://doi.org/10.1038/NMAT3051} {\bibfield  {journal} {\bibinfo  {journal}
  {NATURE MATERIALS}\ }\textbf {\bibinfo {volume} {10}},\ \bibinfo {pages}
  {521} (\bibinfo {year} {2011})}\BibitemShut {NoStop}%
\bibitem [{\citenamefont {Zhang}\ \emph {et~al.}(2021)\citenamefont {Zhang},
  \citenamefont {Gao}, \citenamefont {Xie}, \citenamefont {Po},\ and\
  \citenamefont {Law}}]{zhang2021higherorder}%
  \BibitemOpen
  \bibfield  {author} {\bibinfo {author} {\bibfnamefont {C.-P.}\ \bibnamefont
  {Zhang}}, \bibinfo {author} {\bibfnamefont {X.-J.}\ \bibnamefont {Gao}},
  \bibinfo {author} {\bibfnamefont {Y.-M.}\ \bibnamefont {Xie}}, \bibinfo
  {author} {\bibfnamefont {H.~C.}\ \bibnamefont {Po}},\ and\ \bibinfo {author}
  {\bibfnamefont {K.~T.}\ \bibnamefont {Law}},\ }\href@noop {} {\bibinfo
  {title} {Higher-order nonlinear anomalous hall effects induced by berry
  curvature multipoles}} (\bibinfo {year} {2021}),\ \Eprint
  {https://arxiv.org/abs/2012.15628} {arXiv:2012.15628 [cond-mat.mes-hall]}
  \BibitemShut {NoStop}%
\bibitem [{\citenamefont {Deyo}\ \emph {et~al.}(2009)\citenamefont {Deyo},
  \citenamefont {Golub}, \citenamefont {Ivchenko},\ and\ \citenamefont
  {Spivak}}]{deyo2009semiclassical}%
  \BibitemOpen
  \bibfield  {author} {\bibinfo {author} {\bibfnamefont {E.}~\bibnamefont
  {Deyo}}, \bibinfo {author} {\bibfnamefont {L.~E.}\ \bibnamefont {Golub}},
  \bibinfo {author} {\bibfnamefont {E.~L.}\ \bibnamefont {Ivchenko}},\ and\
  \bibinfo {author} {\bibfnamefont {B.}~\bibnamefont {Spivak}},\ }\href@noop {}
  {\bibinfo {title} {Semiclassical theory of the photogalvanic effect in
  non-centrosymmetric systems}} (\bibinfo {year} {2009}),\ \Eprint
  {https://arxiv.org/abs/0904.1917} {arXiv:0904.1917 [cond-mat.mes-hall]}
  \BibitemShut {NoStop}%
\bibitem [{\citenamefont {Tsirkin}\ and\ \citenamefont
  {Souza}(2021)}]{tsirkin2021separation}%
  \BibitemOpen
  \bibfield  {author} {\bibinfo {author} {\bibfnamefont {S.~S.}\ \bibnamefont
  {Tsirkin}}\ and\ \bibinfo {author} {\bibfnamefont {I.}~\bibnamefont
  {Souza}},\ }\href@noop {} {\bibinfo {title} {On the separation of hall and
  ohmic nonlinear responses}} (\bibinfo {year} {2021}),\ \Eprint
  {https://arxiv.org/abs/2106.06522} {arXiv:2106.06522 [cond-mat.mtrl-sci]}
  \BibitemShut {NoStop}%
\bibitem [{\citenamefont {Zhang}\ \emph {et~al.}(2018)\citenamefont {Zhang},
  \citenamefont {Ishizuka}, \citenamefont {van~den Brink}, \citenamefont
  {Felser}, \citenamefont {Yan},\ and\ \citenamefont
  {Nagaosa}}]{PhysRevB.97.241118}%
  \BibitemOpen
  \bibfield  {author} {\bibinfo {author} {\bibfnamefont {Y.}~\bibnamefont
  {Zhang}}, \bibinfo {author} {\bibfnamefont {H.}~\bibnamefont {Ishizuka}},
  \bibinfo {author} {\bibfnamefont {J.}~\bibnamefont {van~den Brink}}, \bibinfo
  {author} {\bibfnamefont {C.}~\bibnamefont {Felser}}, \bibinfo {author}
  {\bibfnamefont {B.}~\bibnamefont {Yan}},\ and\ \bibinfo {author}
  {\bibfnamefont {N.}~\bibnamefont {Nagaosa}},\ }\bibfield  {title} {\bibinfo
  {title} {Photogalvanic effect in weyl semimetals from first principles},\
  }\href {https://doi.org/10.1103/PhysRevB.97.241118} {\bibfield  {journal}
  {\bibinfo  {journal} {Phys. Rev. B}\ }\textbf {\bibinfo {volume} {97}},\
  \bibinfo {pages} {241118(R)} (\bibinfo {year} {2018})}\BibitemShut {NoStop}%
\bibitem [{\citenamefont {Le}\ \emph {et~al.}(2020)\citenamefont {Le},
  \citenamefont {Zhang}, \citenamefont {Felser},\ and\ \citenamefont
  {Sun}}]{PhysRevB.102.121111}%
  \BibitemOpen
  \bibfield  {author} {\bibinfo {author} {\bibfnamefont {C.}~\bibnamefont
  {Le}}, \bibinfo {author} {\bibfnamefont {Y.}~\bibnamefont {Zhang}}, \bibinfo
  {author} {\bibfnamefont {C.}~\bibnamefont {Felser}},\ and\ \bibinfo {author}
  {\bibfnamefont {Y.}~\bibnamefont {Sun}},\ }\bibfield  {title} {\bibinfo
  {title} {Ab initio study of quantized circular photogalvanic effect in chiral
  multifold semimetals},\ }\href {https://doi.org/10.1103/PhysRevB.102.121111}
  {\bibfield  {journal} {\bibinfo  {journal} {Phys. Rev. B}\ }\textbf {\bibinfo
  {volume} {102}},\ \bibinfo {pages} {121111(R)} (\bibinfo {year}
  {2020})}\BibitemShut {NoStop}%
\end{thebibliography}%

\appendix
\section{Expressions for the Green functions and the self energies in
  the Moyal-Keldysh approach}
\label{sec_appendix}

In Eq.~\eqref{eq_efiefi_lesser_all}
\bege\label{eq_glestwo_ee}
\glestwo_{E_i,E_j}(\mathcal{E})=
\left[
\gadv_{E_i,E_j}(\mathcal{E})-\gret_{E_i,E_j}(\mathcal{E})
\right]
\ee
determines the contribution that is
proportional to the Fermi function $f(\mathcal{E})$.
Here, the retarded function $\gret_{E_i,E_j}$ 
is given by
\bege\label{eq_retarded_ee}
\begin{aligned}
&\gret_{E_i,E_j}=
\gret_{0}\Biggl[
\sigmaret_{E_i,E_j}
\gret_{0}+
\sigmaret_{E_i}
\gret_{E_j}+
\sigmaret_{E_j}
\gret_{E_i}+\\
&-\frac{i}{2}
\partial_{\pi_i}\sigmaret_{E_j}
\partial_{\pi_0}\gret_{0}
+\frac{i}{2}
\partial_{\pi_i}(\gret_{0})^{-1}
\partial_{\pi_0}\gret_{E_j}\\
&+\frac{i}{2}
\partial_{\pi^{0}}
\sigmaret_{E_j}
\partial_{\pi_i}
\gret_{0}
-\frac{i}{2}
\partial_{\pi^{0}}
(\gret_{0})^{-1}
\partial_{\pi^i}
\gret_{E_j}\\
&-\frac{i}{2}
\partial_{\pi_j}
\sigmaret_{E_i}
\partial_{\pi_0}
\gret_{0}
+\frac{i}{2}
\partial_{\pi_j}
(\gret_{0})^{-1}
\partial_{\pi_0}
\gret_{E_i}\\
&+\frac{i}{2}
\partial_{\pi_0}
\sigmaret_{E_i}
\partial_{\pi_j}
\gret_{0}
-\frac{i}{2}
\partial_{\pi_0}
(\gret_{0})^{-1}
\partial_{\pi_j}
\gret_{E_i}\\
&+\frac{1}{4}
\partial_{\pi_i}\partial_{\pi_j}
(\gret_{0})^{-1}
\partial_{\pi_0}\partial_{\pi_0}
\gret_{0}\\
&-\frac{1}{4}
\partial_{\pi_0}\partial_{\pi_j}
(\gret_{0})^{-1}
\partial_{\pi_0}\partial_{\pi_i}
\gret_{0}\\
&-\frac{1}{4}
\partial_{\pi_0}\partial_{\pi_i}
(\gret_{0})^{-1}
\partial_{\pi_0}\partial_{\pi_j}
\gret_{0}\\
&+\frac{1}{4}
\partial_{\pi_0}\partial_{\pi_0}
(\gret_{0})^{-1}
\partial_{\pi_j}\partial_{\pi_i}
\gret_{0}\Biggr]\\
\end{aligned}
\ee
and 
$\gadv_{E_i,E_j}=[\gret_{E_i,E_j}]^{\dagger}$,
where
\bege\label{eq_green_efi}
\begin{aligned}
G^{\rm R}_{E_{i}}
&=G^{\rm R}_{0}\Biggl[\sigmaret_{E_{i}}+\frac{i}{2}\frac{c}{\hbar}
\Bigl[
\frac{\partial}{\partial \mathcal{E}}
(G^{\rm R}_{0})^{-1}
G^{\rm R}_{0}
\frac{\partial}{\partial k_i}
(G^{\rm R}_{0})^{-1}\\
&-
\frac{\partial}{\partial k_i}
(G^{\rm R}_{0})^{-1}
G^{\rm R}_{0}
\frac{\partial}{\partial \mathcal{E}}
(G^{\rm R}_{0})^{-1}
\Bigr]
\Biggr]G^{\rm R}_{0}.\\
\end{aligned}
\ee

Additionally,
\bege\label{eq_glesone_ee}
\begin{aligned}
&\glesone_{E_i,E_j}=\gret_{0}\Biggl[
\sigmalesone_{E_i,E_j}
\gadv_{0}+
\sigmalesone_{E_i}
\gadv_{E_j}+
\sigmalesone_{E_j}
\gadv_{E_i}+
\\
&-\frac{i}{2}
\partial_{\pi_i}\sigmaret_{E_j}
[
\gadv_{0}-\gret_{0}
]
+\frac{i}{2}
\partial_{\pi_i}(\gret_{0})^{-1}
\partial_{\pi_0}\glesone_{E_j}\\
&+\frac{i}{2}
\partial_{\pi_i}(\gret_{0})^{-1}
\glestwo_{E_j}
\\
&+\frac{i}{2}
\sigmalestwo_{E_j}
\partial_{\pi_i}
\gadv_{0}
+\frac{i}{2}
\partial_{\pi^{0}}
\sigmalesone_{E_j}
\partial_{\pi_i}
\gadv_{0}
\\
&-\frac{i}{2}
[
(\gadv_{0})^{-1}
-
(\gret_{0})^{-1}
]
\partial_{\pi^i}
\gadv_{E_j}
-\frac{i}{2}
\partial_{\pi^{0}}
(\gret)^{-1}
\partial_{\pi^i}
\glesone_{E_j}\\
&-\frac{i}{2}
\partial_{\pi_j}
\sigmaret_{E_i}
[
\gadv_{0}
-
\gret_{0}
]
+\frac{i}{2}
\partial_{\pi_j}
(\gret_{0})^{-1}
\partial_{\pi_0}
\glesone_{E_i}\\
&+\frac{i}{2}
\partial_{\pi_j}
(\gret_{0})^{-1}
\glestwo_{E_i}\\
&+\frac{i}{2}
\partial_{\pi_0}
\sigmalesone_{E_i}
\partial_{\pi_j}
\gadv_{0}
+\frac{i}{2}
\sigmalestwo_{E_i}
\partial_{\pi_j}
\gadv_{0}
\\
&-\frac{i}{2}
[
(\gadv_{0})^{-1}-
(\gret_{0})^{-1}
]
\partial_{\pi_j}
\gadv_{E_i}-\frac{i}{2}
\partial_{\pi_0}
(\gret_{0})^{-1}
\partial_{\pi_j}
\glesone_{E_i}\\
&+\frac{1}{2}
\partial_{\pi_i}\partial_{\pi_j}
(\gret_{0})^{-1}
[
\partial_{\pi_0}
\gadv_{0}
-
\partial_{\pi_0}
\gret_{0}
]\\
&-\frac{1}{4}
\partial_{\pi_0}\partial_{\pi_j}
(\gret_{0})^{-1}
[\partial_{\pi_i}\gadv_{0}
-
\partial_{\pi_i}\gret_{0}]
\\
&-\frac{1}{4}
[
\partial_{\pi_j}
(\gadv_{0})^{-1}
-
\partial_{\pi_j}
(\gret_{0})^{-1}
]
\partial_{\pi_0}\partial_{\pi_i}
\gadv_{0}\\
&-\frac{1}{4}
\partial_{\pi_0}\partial_{\pi_i}
(\gret_{0})^{-1}
[
\partial_{\pi_j}
\gadv_{0}
-
\partial_{\pi_j}
\gret_{0}
]\\
&-\frac{1}{4}
[\partial_{\pi_i}
(\gadv_{0})^{-1}
-
\partial_{\pi_i}
(\gret_{0})^{-1}
]
\partial_{\pi_0}\partial_{\pi_j}
\gadv_{0}\\
&+\frac{1}{2}
[
\partial_{\pi_0}
(\gadv_{0})^{-1}
-
\partial_{\pi_0}
(\gret_{0})^{-1}
]
\partial_{\pi_j}\partial_{\pi_i}
\gadv_{0}\Biggr]\\
\end{aligned}
\ee
determines the contribution that is
proportional to the energy-derivative
of the Fermi
function $f'(\mathcal{E})=\partial f/\partial\mathcal{E}$,
and
\bege\label{eq_glesthree_ee}
\begin{aligned}
&\glesthree_{E_i,E_j}=
\gret_{0}\Biggl[
\sigmalesthree_{E_i,E_j}
\gadv_{0}
+\frac{i}{2}
\partial_{\pi_i}(\gret_{0})^{-1}
\glesone_{E_j}\\
&+\frac{i}{2}
\sigmalesone_{E_j}
\partial_{\pi_i}
\gadv_{0}
+\frac{i}{2}
\partial_{\pi_j}
(\gret_{0})^{-1}
\glesone_{E_i}\\
&+\frac{i}{2}
\sigmalesone_{E_i}
\partial_{\pi_j}
\gadv_{0}
+\frac{1}{4}
\partial_{\pi_i}\partial_{\pi_j}
(\gret_{0})^{-1}
[
\gadv_{0}
-
\gret_{0}
]
\\
&+\frac{1}{4}
[
(\gadv_{0})^{-1}
-
(\gret_{0})^{-1}
]
\partial_{\pi_j}\partial_{\pi_i}
\gadv_{0}\Biggr]\\
\end{aligned}
\ee
determines the contribution that is
proportional to the second energy-derivative
of the Fermi
function $f''(\mathcal{E})=\partial^2 f/\partial\mathcal{E}^2$.
Here,
\bege\label{eq_gles_efi}
\begin{aligned}
\gles_{E_i,I}&=\gret_{0}
\sigmales_{E_i,I}
\gadv_{0}\\
&-\frac{i}{2}\frac{c}{\hbar}\Biggl[
\gret_{0}
\left[
\frac{\partial H}{\partial k_i}
+
\frac{\partial \sigmaret_0}{\partial k_i}
\right]
\left[
\gadv_{0}-\gret_{0}
\right]
\\
&-
\left[
\gadv_{0}-\gret_{0}
\right]
\left[
\frac{\partial H}{\partial k_i}
+
\frac{\partial \sigmaadv_0}{\partial k_i}
\right]\gadv_{0}
\Biggr].\\
\end{aligned}
\ee

In the Gaussian disorder approximation we
determine the self energies from the equations~\cite{theory_noneq_const_emf}
\bege\label{eq_app_sigma}
\Sigma^{\eta,\#}=
\mathcal{V}\intkspa G^{\eta,\#},
\ee
\bege\label{eq_app_sigma_efi}
\Sigma^{\eta,\#}_{E_{i}}=
\mathcal{V}\intkspa G^{\eta,\#}_{E_{i}},
\ee
and
\bege\label{eq_sigma_efiefi}
\Sigma^{\eta,\#}_{E_{i},E_{j}}=
\mathcal{V}\intkspa G^{\eta,\#}_{E_{i},E_{j}},
\ee
where 
$\mathcal{V}$ quantifies the strength of the disorder scattering and
$\eta={\rm R, A,} <$.
If $\eta\ne<$ we leave $\#$ blank, otherwise $\#={\rm I, II, III}$. 
Since the
Green functions $G^{\eta,\#}$,
$G^{\eta,\#}_{E_{i}}$ 
and $G^{\eta,\#}_{E_{i},E_{j}}$
depend on the self-energies
$\Sigma^{\eta,\#}$,
$\Sigma^{\eta,\#}_{E_{i}}$,
and $\Sigma^{\eta,\#}_{E_{i},E_{j}}$,
the equations Eq.~\eqref{eq_app_sigma},
Eq.~\eqref{eq_app_sigma_efi}
and Eq.~\eqref{eq_sigma_efiefi}
need to be solved self-consistently.
It is straightforward to extend these expressions into the $T$-matrix
approximation~\cite{theory_noneq_const_emf}.
\end{document}